\documentclass[12pt]{article}
\usepackage{amsmath}\usepackage{float}
\usepackage[psamsfonts]{amssymb}
\usepackage{gensymb}
\usepackage{centernot}
\usepackage{graphics,epsfig}
\usepackage{enumitem}
\usepackage[margin=1in]{geometry} 
\usepackage{graphicx}
\graphicspath{ {Images/} }
\usepackage{subcaption}
\usepackage{caption}
\usepackage{multicol}
\usepackage{hhline}
\usepackage{xcolor}
\usepackage{sectsty}
\usepackage{multirow}

\usepackage{mathtools,url}
\usepackage[utf8]{inputenc}
\usepackage[T1]{fontenc}

\usepackage{csvsimple,longtable, tabularx}

\usepackage{array}
\newcommand{\PreserveBackslash}[1]{\let\temp=\\#1\let\\=\temp}
\newcolumntype{C}[1]{>{\PreserveBackslash\centering}p{#1}}

\sectionfont{\fontsize{14}{15}\selectfont}
\subsectionfont{\fontsize{12}{15}\selectfont}

\usepackage{ulem}

\begin{document}

\author{Adrienne C. Kinney$^{1,\ast}$, Sean Current$^{5}$, Joceline Lega$^{2,3,4}$
\\
\normalsize{$^{1}$ Interdisciplinary Program in Applied Mathematics}\\
\normalsize{$^{2}$ Department of Mathematics}\\
\normalsize{$^{3}$ Department of Epidemiology and Biostatistics}\\
\normalsize{$^{4}$ BIO5 Institute}\\
\normalsize{The University of Arizona, Tucson, AZ, 85721, USA}\\
\normalsize{$^{5}$ Department of Computer Science and Engineering}\\
\normalsize{The Ohio State University, Columbus, OH, 43202, USA}\\
\normalsize{$^\ast$To whom correspondence should be addressed;}\\
\normalsize{E-mail:  akinney1@math.arizona.edu.}}
\date{\today}
\title{\textit{Aedes-AI:} Neural Network Models of Mosquito Abundance}



\maketitle
\begin{abstract}
We present artificial neural networks as a feasible replacement for a mechanistic model of mosquito abundance. We develop a feed-forward neural network, a long short-term memory recurrent neural network, and a gated recurrent unit network. We evaluate the networks in their ability to replicate the spatiotemporal features of mosquito populations predicted by the mechanistic model, and discuss how augmenting the training data with time series that emphasize specific dynamical behaviors affects model performance. We conclude with an outlook on how such equation-free models may facilitate vector control or the estimation of disease risk at arbitrary spatial scales.
\end{abstract}

\section{Introduction}

Artificial Neural Networks (ANNs) are ideally suited for modeling nonlinear, complex phenomena, and have either achieved or surpassed human-level performance on tasks involving image classification, anomaly detection, and event extraction \cite{krizhevsky2012imagenet, kwon2019survey, nguyen2016joint, schmidhuber2015deep}. This manuscript assesses the feasibility of developing ANNs that provide estimates of mosquito abundance directly from weather data. To this end, we train, validate, and test different types of neural networks on simulated datasets. The training data consist of years of representative daily weather time series for various locations in the US obtained from the Multivariate Adaptive Constructed Analogs (MACA) datasets \cite{MACA}, and of corresponding daily mosquito abundance predictions. The latter are estimated from a mechanistic model, the Mosquito Landscape Simulation (MoLS) \cite{lega2017aedes}, which uses the MACA data as input. Our goal is to train ANNs that learn how MoLS estimates mosquito abundance from weather time series. We introduce various metrics that allow us to compare the ANN predictions to those of MoLS, including criteria relevant to public health (e.g. mosquito season length and timing of peak mosquito abundance) and rank the proposed ANNs based on a composite score derived from these metrics. We conclude that artificial neural networks are able to quickly and accurately reproduce MoLS daily mean mosquito abundance estimates, provided they are given current and past weather data over a specific window of time; in the present case, we use windows of length 90 days, consisting of the day each abundance estimate is made and the previous 89 days. 

Vector-borne diseases infect hundreds of millions of people annually, disproportionately impacting impoverished communities in tropical areas \cite{world2017global}. The past two decades have seen a surge of outbreaks associated with the vector \textit{Aedes aegypti}, including a 2004 dengue outbreak in Singapore \cite{ong2007fatal}, a 2013-2014 chikungunya outbreak in the Americas \cite{Cauchemez2014local}, and the 2014-2015 Zika outbreak in Latin America \cite{heukelbach2016zika}. These outbreaks exemplify the public health risk of arboviral diseases associated with severe clinical symptoms, but also demonstrate the importance of local vector control efforts to mitigate impact on affected communities \cite{ooi2006dengue,barrera2017impact,world2016mosquito}. Because vector-borne disease outbreaks require sufficiently high vector populations \cite{ryan2006,guo2014,barrera2017impact}, the ability to predict vector abundance is a central component of assessing disease risk. Forecasting outbreaks is, however, further complicated by global interconnectedness \cite{kraemer2015global} and climate change \cite{rocklov2020climate,kamal2018mapping} -- factors shown to introduce vectors into previously uninhabited areas and increase the viable range of vectors, respectively.

MoLS \cite{lega2017aedes} is a mechanistic stochastic model that estimates \textit{Aedes aegypti} abundance from weather time series (temperature, precipitation, and relative humidity). It is parametrized from information previously published in the literature and its output is proportional to expected  daily mosquito numbers. It was shown in \cite{lega2017aedes} that when given accurate local weather data, MoLS is able to reproduce \textit{Aedes aegypti} abundance trends observed in surveillance traps in Puerto Rico. Because its parameters are fixed and thus not location-dependent, MoLS is in principle able to estimate weather-related abundance anywhere, provided local weather information is available, and assuming the biological properties of local \textit{Aedes aegypti} mosquitoes are sufficiently well captured by the model parameters. Scaling this up to a large number of locations is however resource intensive. Specifically, using MoLS to generate daily abundance predictions associated with 10 years of weather data (from 01/01/2011 to 12/31/2020) takes about 10 minutes per location (including file reading and writing time) on one core of a high power computer (HPC) AMD Zen2 node. For county-scale resolution in the US, computing 10 years worth of daily abundance estimates therefore requires about $3000/6 = 500$ HPC hours. The advantage of a HPC is of course that predictions for many locations can be run in parallel, reducing the user's wait time by a factor equal to the total number of cores available.

A faster, ANN-based alternative to MoLS is appealing since it would reduce computational time even further, thereby allowing for a corresponding increase in the spatial resolution of abundance estimates. Like MoLS predictions, the ANN output would reflect an arbitrary carrying capacity, and would need to be scaled to average abundance inferred from surveillance information, in order to account for local conditions (e.g. due to the presence of a variable number of water containers where mosquitoes can lay eggs). Of course, because the output of MoLS or any of its ANN replacements is dependent on input weather data, the accuracy of such estimates is limited by the reliability of the available weather information. Moreover, human influences on mosquito populations (see e.g. \cite{Hemme10,Brown14,wilke20} and references therein), other than consequences of climate change on weather data, are not taken into account in these models. Nevertheless, given that weather plays an important role in \textit{Aedes aegypti} numbers \cite{Halstead08,Valdez18,Benitez21}, quickly producing weather-based abundance estimates of this disease vector is important.

The ability to replace a complex mechanistic model by an ANN that is faster and can easily be scaled up opens a new range of applications for these models. However, because artificial neural network predictions often lose accuracy in ``unfamiliar'' situations, the training data may need to place special emphasis on specific dynamic behaviors that are deemed important by the modeler. Estimating what type and what fraction of additional information is necessary to improve performance plays an important role in the development of ANN-based models. It is these questions that have motivated the work presented in this manuscript. Although they are addressed in the specific context of {\it Aedes aegypti} abundance, the approaches discussed here are general, and can be extended to other mechanistic models of vector abundance, such as for instance {\it DyMSim} \cite{morin2010modeled}, a model for the abundance of {\it Culex} species.

\section{Methods}\label{sec:Methods}

\subsection{Training and Input Data}\label{sec:training_input_data}

The neural network models discussed in this article have the same input data as MoLS. These consist of daily time series of maximum temperature, minimum temperature, precipitation, and average relative humidity. In MoLS, this information is used to calculate \textit{Aedes aegypti} development, death, and reproductive rates, simulate daily mosquito abundance, and estimate the daily expectation of the number of gravid females \cite{lega2017aedes}. The ANNs work differently: at any given location, a trained ANN converts weather time series for a fixed number of consecutive days into a single number, which is the estimated gravid female abundance at the given location on the last day of the given time series. Such an ANN is trained on a dataset consisting of weather time series and associated MoLS estimates.

\subsubsection{Weather Input Data}

\begin{figure}[ht]
    \centering
    \includegraphics[width=\textwidth]{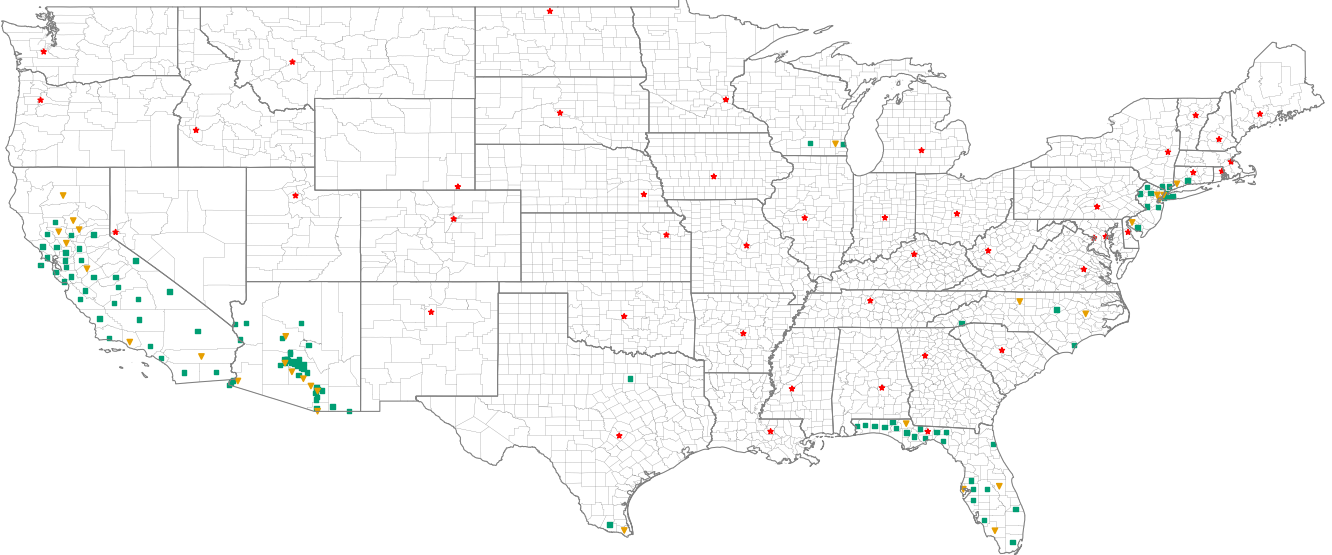}
    \caption{\label{fig:map} Map of the contiguous United States showing the locations used for training and validation (green squares), and testing (orange triangles). The locations used in the \textit{Capital Cities} dataset are indicated by red stars. See Tables \ref{tab:training_locs} and \ref{tab:testing_locs} in Appendix \ref{app:D} for the names of the locations in the principal dataset.}
\end{figure}

We obtain daily weather time series from the Multivariate  Adaptive  Constructed  Analogs  (MACA) datasets website \cite{MACA}. To train, validate, and test the models, we define a principal dataset consisting of daily data for the years 2012-2020 at 144 locations in 9 states: Arizona, California, Connecticut, Florida, New Jersey, New York, North Carolina, Texas, and Wisconsin. States other than Arizona are chosen because of their participation in the 2019 CDC \textit{Aedes} Challenge \cite{CDC_AC} and the locations in this study are the centroids of counties that provided data for the challenge. For Arizona, we use MoLS predictions for the 50 most populated cities in the state. Together, these locations exemplify varying mosquito population patterns associated with different climates: hot and dry summers, hot and humid summers, cold winters, etc. We define a second dataset, called \textit{Capital Cities}, to assess the performance of the trained ANNs across the contiguous US, in previously unseen locations (see Appendix \ref{app:CC}). To this end, we downloaded  the 2012-2020 MACA time series for all capital cities that are not situated in counties included in the principal dataset. Figure \ref{fig:map} shows the locations whose time series we use for training, validation, and testing of the ANN models.

The weather time series used as input are noisy at the daily scale (with a correlation length of about 2 to 4 weeks for temperature, depending on location, and of about 6 days for relative humidity; data not shown) but exhibit seasonal patterns. Figure \ref{fig:Weather_AZ} of Appendix \ref{app:A1} illustrates the temporal dynamics of daily average temperature, precipitation, and relative humidity in Phoenix, AZ, as well as the dependence of these quantities on two climate models (US GFDL-ESM2M and Canada CanESM2). Figure \ref{fig:location_correlations} of Appendix \ref{app:A1} (left two panels) shows the correlation of the weather between locations used for training and those used for testing. Choosing a range of geographical locations ensures the ANN models are tested on samples of varying weather trends.

\subsubsection{Mosquito Abundance Input Data}

MoLS \cite{lega2017aedes} is initiated with a specified number of \textit{Aedes aegypti} eggs, and the simulation follows the life cycle of each egg ``laid'' in the system. An egg must survive through five immature stages before emerging as a fertile adult. At each stage in the life cycle, MoLS uses environmental and entomological features to simulate the lifespan of the mosquitoes, including temperature-dependent development rates and gonotrophic cycles, daily survival rates that depend on temperature and relative humidity, precipitation-dependent egg hatching, and carrying capacities estimated from water levels in simulated containers. MoLS takes about 10-12 weeks to ramp up, after which the simulated pool of mosquitoes (eggs, larvae, pupae, and adults) becomes representative of the weather data and local carrying capacity. Although MoLS output includes information on all of the life stages of a mosquito population, its default output is daily scaled gravid female abundance. This allows for direct comparison with surveillance data, which are often collected in gravid mosquito traps. More information about MoLS, including a comparison of its gravid female mosquito predictions against trap data for four neighborhoods in Puerto Rico may be found in \cite{lega2017aedes}. 

In contrast to weather data, MoLS time series show little noise because they represent abundance expectation, smoothed over a two-week window. To illustrate how MoLS responds to changes in its input time series, its predictions for Phoenix, AZ, associated with weather data from two climate models are shown in Figure \ref{fig:MoLS_AZ} of Appendix \ref{app:A1}. The correlation of its estimates between testing and training locations is presented in the right panel of Figure \ref{fig:location_correlations}.

\subsubsection{Weather Input Sample Length}

MoLS keeps track of the number of eggs laid by adult female mosquitoes over many generations and, as a consequence, ANNs cannot be expected to reproduce MoLS results with only one day of weather data. Instead, they are provided with weather information over a time window $[t_0 - \Delta + 1, t_0]$ of fixed length $\Delta$ days, in order to estimate abundance on day $t_0$. Because MoLS takes about 10-12 weeks to ramp up, we expect $\Delta$ to be of comparable length, i.e. 90 days. Although large enough windows are needed for good performance, there is a trade-off between larger values of $\Delta$ and accuracy. Longer windows require users to provide reliable weather data over longer periods of time and increase computational cost. Moreover, windows that are too long may teach the ANNs to rely too much on what happened during the previous mosquito season. The models discussed in this article use $\Delta = 90$ days and are able to reproduce MoLS output with high skill. For comparison, we provide an example of an ANN trained with $\Delta = 120$ days in Appendix \ref{App:other_models}. We note that the average lifetime of a mosquito is estimated to be 30 days (about two weeks in immature stages \cite{lega2017aedes} and two weeks in the adult stage \cite{maciel-de-freitas2007}). Getting good results with $\Delta = $ 90 days suggests that 3 times the average individual lifespan is sufficient to capture any correlation between current and future population trends.

\subsection{Models}

\subsubsection{Baseline Model}

We utilize a simple linear regression model (LR) optimized with gradient descent as a baseline model for comparison. The linear regression model is trained on the same training subset as the ANNs and its weights are found using an Adam optimizer \cite{kingma2014adam} with learning rate $\alpha=0.0001$. Note that the LR model can output negative values; this is fixed in post processing by taking the maximum of the output and zero.

\subsubsection{Neural Network Models}\label{sec:Models}
\begin{figure}[ht]
    \centering
    \includegraphics[scale=0.3]{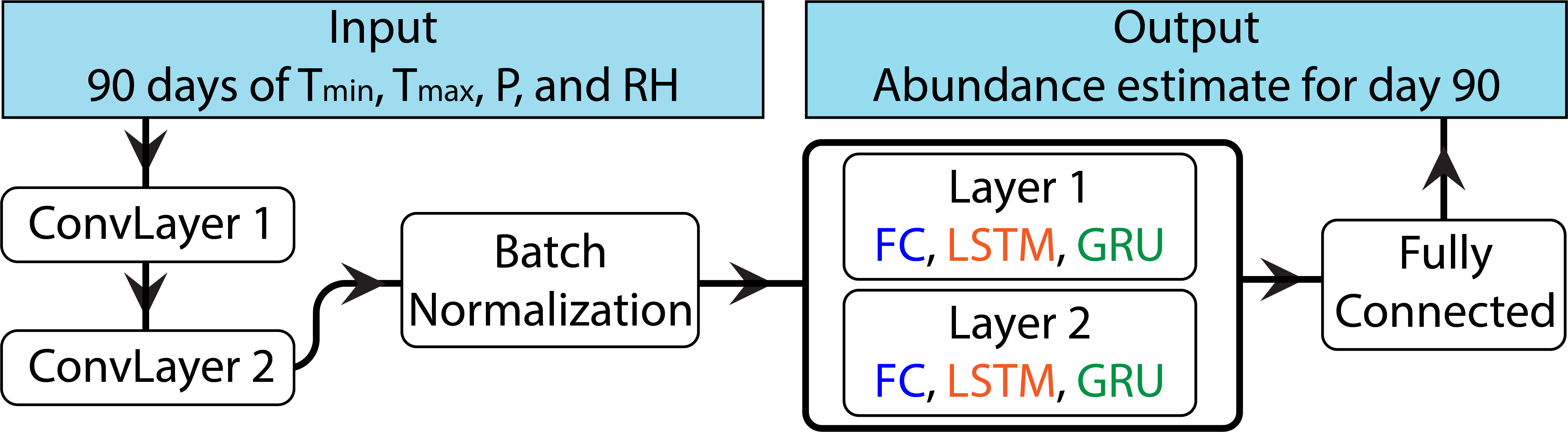}
    \caption{\label{fig:arch} Architecture of the models used in this study.}
\end{figure}
We define three neural network models: a feed-forward neural network, a long short-term memory neural network, and a gated recurrent unit neural network. A schematic is provided in Figure \ref{fig:arch}. The layers in the thick-edged box are model dependent: fully-connected (FC) for Model 1, long short-term memory recurrent units (LSTM) for Model 2, and gated recurrent units (GRU) for Model 3, all of which we describe below. Each model begins with two convolution layers with 64 units each, a kernel size of 3 with no padding, and a stride of 1. Both layers use rectified linear unit (ReLU) activation and are immediately followed by a batch normalization layer as a form of regularization. Batch normalization scales layer outputs according to a learned mean and standard deviation to reduce overfitting and improve generalizability. We considered dropout methods as an alternate means to reduce overfitting, but removed them after they demonstrated no noticeable improvement on the validation set. The number of trainable parameters for each model is shown in Table \ref{tab:train params}. Additionally, the reader is referred to Appendix \ref{app:A} for details on the layers we use in the models and to \cite{Goodfellow-et-al-2016} for a more thorough discussion of neural networks. The loss function and optimization are discussed in \S\ref{sec:Hyperparameter Selection}.

\begin{table}[ht]
    \centering
    \begin{tabular}{|c|c|}
    \hline
         \textbf{Model} & \textbf{Trainable Parameters}  \\
         \hline
         FF & 369,857\\
         \hline
         LSTM & 79,425\\
         \hline
         GRU & 63,297 \\
    \hline
    \end{tabular}
    \caption{Number of trainable parameters for each of the neural network models.}
    \label{tab:train params}
\end{table}

\noindent {\bf Model 1: Feed-forward convolutional neural network (FF).}
The feed-forward network flattens the batch normalization output before applying two fully connected layers and an output layer. The fully connected layers have 64 units and ReLU activation. The output layer is a single unit with ReLU activation and $\ell_2$ regularization to reduce over-fitting on the training data. The $\ell_2$ regularization on the output augments the loss function with the $\ell_2$ norm of the output weights, $||w||_{2}$, which must be minimized alongside the mean square error ($MSE$) loss. This penalizes large weight terms, requiring the model to utilize multiple features in its decision making, avoiding over-fitting as a result.

\bigskip
\noindent {\bf Model 2: Long short-term memory recurrent neural network (LSTM).}
Our second model architecture is an LSTM, chosen to exploit the ``memory'' feature of recurrent neural networks. LSTM units include gates that selectively allow information to propagate forward, thereby making it possible for previous information to directly influence the model's behavior. Such a feature is relevant for abundance predictions since previous weather patterns impact current populations. For example, significant heat or cold decreases the viability of offspring, limiting future abundance. Moreover, \textit{Aedes aegypti} eggs are known to be resistant to desiccation: long droughts do not necessarily cause a decrease in viable eggs, which can later hatch when rainfall creates new habitat (see for instance \cite{brown2016} and references therein). All of these features are taken into account in MoLS and are thus expected to be captured by the ANNs. The architecture of the LSTM model replaces the two fully connected layers of Model 1 with LSTM layers (Figure \ref{fig:arch}), each with 64 units and {\tt tanh} activation. 

\bigskip
\noindent {\bf Model 3: Gated recurrent unit recurrent neural network (GRU).}
The final model architecture we consider is a GRU, chosen to leverage the benefits of the LSTM model while reducing the number of associated parameters. GRUs, like LSTMs, feature a gated unit to selectively allow information to propagate forward. However, the GRU unit is simpler than a LSTM unit (see Table \ref{tab:train params}), which reduces training time and the computational cost of using the model to generate predictions.
The GRU architecture is identical to the LSTM architecture, except the LSTM layers are replaced by two GRU layers, each with 64 units and {\tt tanh} activation (Figure \ref{fig:arch}).

\subsection{Model Training}\label{sec:Model Training}

\subsubsection{Data Processing}\label{sec:Data Processing}

We define subsets of the principal dataset for training, validation, and testing. The training subset contains daily weather data and corresponding MoLS predictions from 2012-2018 for 115 locations, shown in green (squares) in Figure \ref{fig:map}, and we use it to set the weights in the ANNs. The validation subset contains data from 2019-2020 for the same 115 locations, and is used during hyperparmeter selection (\S\ref{sec:Hyperparameter Selection}) to optimize model performance. The testing subset contains the daily weather data from 2012-2020 for the 29 locations not included in the training and validation subsets, shown in orange (triangles) in Figure \ref{fig:map}. The \textit{Capital Cities} dataset contains daily weather data from 2012-2020 for capital cities in the contiguous US that are not in a county used in the principal dataset. The testing data do not include the corresponding MoLS time series, which are subsequently used to evaluate the performance of the optimized models in terms of the metrics defined in Section \ref{sec:New Metrics Section}.

During the training and validation process of each ANN model, we process the input weather data in samples, where the $i$th input sample, $\boldsymbol{x}_i\in\mathbb{R}^{90\times4}$, represents 90 consecutive days of daily observations for the four weather variables (precipitation, maximum temperature, minimum temperature, and relative humidity) at a given location. For each training and validation input sample $\boldsymbol{x}_i$, we define the corresponding output target, $y_i\in\mathbb{R}$, as the gravid female abundance prediction by MoLS for the 90th day of the input sample at the same location. One thousand input samples, and their corresponding output targets, are randomly selected from each location in the training and validation subsets, and randomly shuffled to ensure the model is not dependent on spatiotemporal relationships among successive samples. We scale each sample between 0 and 1 using the global minimum and maximum values of each weather variable for the entire training subset before passing them to one of the ANN models. The resulting scaling factors are considered model parameters and are required for processing future weather samples. All future data, such as validation and testing data, are scaled using the same global minimum and maximum values as the training data. This ensures the data maintain the same relative scale across locations, while removing the differences in scale between temperature, precipitation, and humidity variables. The training samples are used to optimize the loss function and update the model layer parameters, while the validation samples guide hyperparameter selection, described in \S\ref{sec:Hyperparameter Selection}.

After the training and validation process, the learned model weights, as well as the training data extrema, are saved and can be used to make predictions on unseen data. For the testing subset and the \textit{Capital Cities} dataset, we again create input samples $\boldsymbol{x}_i\in\mathbb{R}^{90\times4}$ and use the ANNs to generate abundance estimates on the last day of each testing sample.
For each combination of training, validation, and testing location and year, we create an associated abundance curve by constructing a time series of consecutive daily abundance estimates.

\subsubsection{Loss Function and Hyperparameter Selection}\label{sec:Hyperparameter Selection}

For each ANN model, the model weight parameters are selected during training by minimizing a loss function, defined as the mean squared error ($MSE$) between model output and MoLS predictions: $\sum_{i=1}^n (\hat{y}_i-y_i)^2/n$ where $n$ is the number of data points, $\hat{y}_i$ is the $ith$ prediction by the ANN model and $y_i$ is the $i$th prediction by MoLS.

Model hyperparameters include the learning rate $\alpha$, first moment decay rate $\beta_1$, and the second moment decay rate $\beta_2$ of the Adam optimizer \cite{kingma2014adam}, as well as the number of units and type of activation function for each layer in the model. Given the extensive search space of hyperparameters for neural network models, it is not feasible to test all possible combinations of values. We construct each model by initializing layers with few units and iteratively increasing the number of units in each layer until either the desired performance is achieved or diminishing returns on validation performance are observed (``diminishing returns'' are defined holistically; in particular, if the increase in the number of weights offers no significant decrease in the validation error after training, the lesser number of weights is used). Activation functions are similarly tested on a layer-by-layer basis. Finally, we test learning rates $\alpha \in \{0.01, 0.001, 0.0001, 0.00001\}$. We choose $\alpha=0.0001$, while the first moment decay rate $\beta_1=0.9$ and second moment decay rate $\beta_2=0.999$ of the Adam optimizer are kept at their default values after changes in their values demonstrated less efficient optimization patterns.

\begin{figure}
    \centering
    \includegraphics[width=0.8\linewidth]{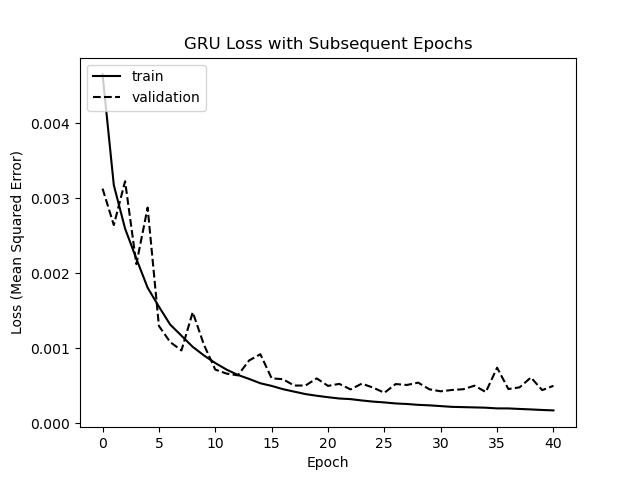}
    \caption{Training loss curve for the GRU model. Training halts due to early stopping after 40 epochs, indicating that the validation loss has reached its minimum value at 25 epochs.}
    \label{fig:training_loss}
\end{figure}

We use a batch size of 64 and train for 100 epochs (enough for models to reach early-stopping convergence criteria; see figure \ref{fig:training_loss}) with an early stopping patience level of 15 epochs. Early stopping prevents over-fitting the training subset by stopping the training process once no improvement is seen in model performance on the validation subset for 15 epochs. Once the early stopping is triggered, the model parameters for the best performing epoch are selected as the learned weights for the model.

\subsection{Data Augmentation}\label{sec:Data Augmentation}

The ``base'' training subset, described in \S\ref{sec:Data Processing}, includes an equal number of input samples from all training locations, but we also define additional training subsets biased towards 1) the double peak pattern observed in hot and dry climates (see Figure \ref{fig:on-off} \textit{right}) and 2) the absence of mosquito populations during colder off-season periods. We test two data augmentation methods: high temperature (HI) oversampling supplements the base training subset with additional samples from on-season periods for double peak locations, while low temperature (LO) oversampling amends the training subset with additional samples from off-season periods from a diverse selection of  locations.

\subsubsection{High Temperature Oversampling (HI)}

The goal of the high temperature oversampling method is to increase the representation of samples featuring the loss of mosquito population due to extreme heat, resulting in the double peak season pattern (Figure \ref{fig:on-off} \textit{right}). We manually identify locations in the original training subset featuring such a pattern (see Table \ref{tab:training_locs} in Appendix \ref{app:D}), and sample 1000 windows of length $\Delta$ at each selected location, which are then incorporated into the training data. Each of these windows is such that its final day lies within the on-season boundaries. As a consequence, the HI training subset includes 1000 randomly chosen input samples from on-season times for each of the locations in Table \ref{tab:training_locs}, in addition to 1000 randomly chosen input samples from each training location.

\subsubsection{Low Temperature Oversampling (LO)}

The second approach is low temperature oversampling. Similar in construction to the high temperature oversampling method, we manually select a diverse set of locations featuring consistent losses of mosquito population due to cold weather, particularly during winter months. We randomly sample 1000 training windows for each of the selected locations, with the final day of the training window lying in the colder, off-season months. Thus, the LO training subset includes 1000 randomly chosen input samples from off-season times for each of the low temperature oversampling locations listed in Table \ref{tab:training_locs}, together with 1000 randomly chosen input samples from each training location.

\subsubsection{Variant Model Training with Augmented Data}

In addition to the HI and LO training subsets, we define the HI LO training subset as the combination of the two; this training subset contains the 1000 randomly chosen input samples from each location, the 1000 randomly chosen HI samples, and the 1000 randomly chosen LO samples. Then we retrain the base models (FF, LSTM, and GRU) using all three new training subsets. We use the same training process as described in \S\ref{sec:Model Training}. Thus, in addition to the three base models we have 9 variant models, named according to the combination of base model and training subset: FF HI, FF LO, FF HI LO, LSTM HI, LSTM LO, LSTM HI LO, GRU HI, GRU LO, and GRU HI LO.

\subsection{Post-Processing and Evaluation}

As mentioned above, for each 90-day input sample, the neural network models output the number of gravid female mosquitoes expected on the 90th day. We evaluate these models by first generating the 2012-2020 abundance curves for each of the testing locations. We then smooth each time series of daily predictions, and assess both the global and seasonal fits compared to the corresponding MoLS abundance curve.

\subsubsection{Data Smoothing}\label{sec:Data Smoothing}

\begin{figure}[h!]
    \centering
    \includegraphics[width=0.8\textwidth]{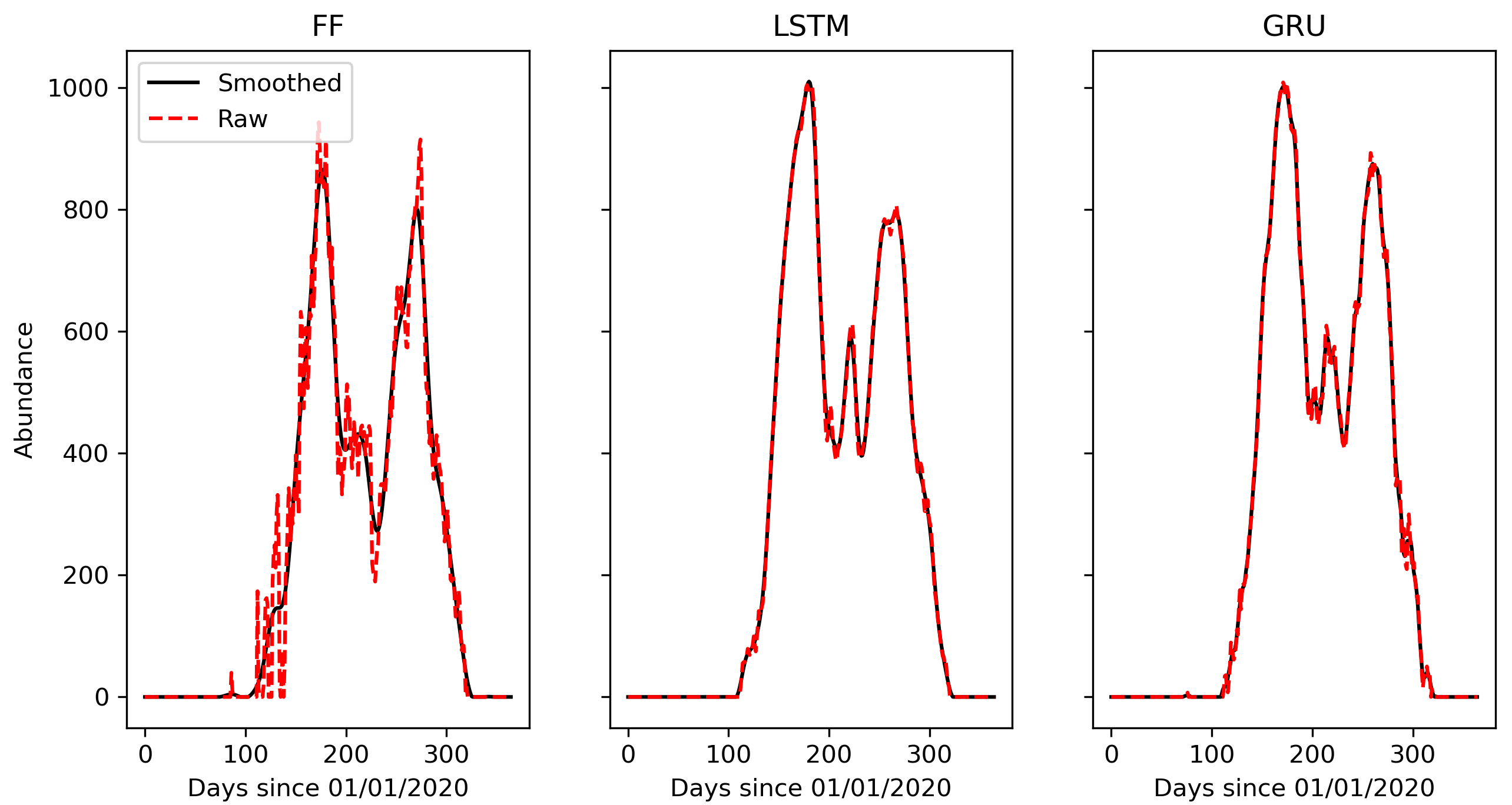}
    \caption{Comparison of the raw and smoothed abundance curves in Avondale, Arizona (2020) for the FF (\textit{left}), LSTM (\textit{center}), and GRU (\textit{right}) models. Data are smoothed according to \S\ref{sec:Data Smoothing}.}
    \label{fig:smoothing}
\end{figure}

Because the output of MoLS is smoothed with two passes of a 15-day moving average filter, we also smooth the ANN time series before evaluating the performance of these models. This is necessary because the weather data and thus the ANN outputs are noisy at the daily scale. We decided not to train the ANNs on the unsmoothed output of MoLS because the latter is an average over a small number (30) of stochastic simulations and the smoothing contributes to producing estimates that represent average abundance. Instead, we expect the ANNs to process daily weather data in a way that produces estimates that fluctuate daily about a time-dependent mean that is as close as possible to MoLS numbers. We use a Savitzky-Golay filter with a window of 11 and polynomial of order 3 to filter the neural network time series. This is an optimal setting that results in having the 11 day auto-correlation of the predictions within $1\%$ of the 11 day auto-correlation of the corresponding MoLS data. The central point of the 3rd order polynomial curve used to fit each 11-day span is returned as the smoothed data point. Any negative values resulting from the smoothing process are set to $0$. Figure \ref{fig:smoothing} shows a comparison of the raw and smoothed abundance curves for Avondale, Arizona in 2020. The reader is referred to \cite{orfanidis1995introduction} for more information on the Savitzky-Golay filter.

\subsubsection{Performance Metrics}
\label{sec:New Metrics Section}

We use a range of metrics to assess the performance of the neural network models. These include four metrics that quantify global fit to the MoLS data, as well as four metrics that focus on timing of abundance peaks and season length. These metrics are then combined into a single score that is used to rank the neural network models.

The global fit metrics are the non-negative coefficient of determination ($R_+^2$), normalized root mean square error ($NRMSE$), relative difference in area under the curve ($Rel.\; AUC\; Diff.$), and Pearson correlation ($r$). Definitions are provided in Appendix \ref{sec:Global Perf Metrics}. While $R_+^2$ and $r$ quantify the fit of the predicted abundance curves, $NRMSE$ and $Rel.\; AUC\; Diff.$ quantify the overall accuracy of the magnitudes of the predicted abundances. High $R_+^2$ and $r$ scores indicate the neural network abundance curves match the shape of the true MoLS curves and low $NRMSE$ and $Rel.\; AUC\; Diff.$ values indicate the magnitudes of the abundance predictions from the neural network models are similar to the corresponding MoLS abundance predictions. These metrics are computed for output samples of varying sizes, such as the entire testing output vector $(n=100,630)$ and the output vector for a specific testing location and year $(n=365)$. We use a normalized $RMSE$ to facilitate comparisons between output samples of different scales (i.e. locations with high mosquito abundance and those with low mosquito abundance) and the non-negative $R_+^2$ to assign a score of $0$ to all low-performing models.

\begin{figure}[ht]
    \centering
    \includegraphics[width=\textwidth]{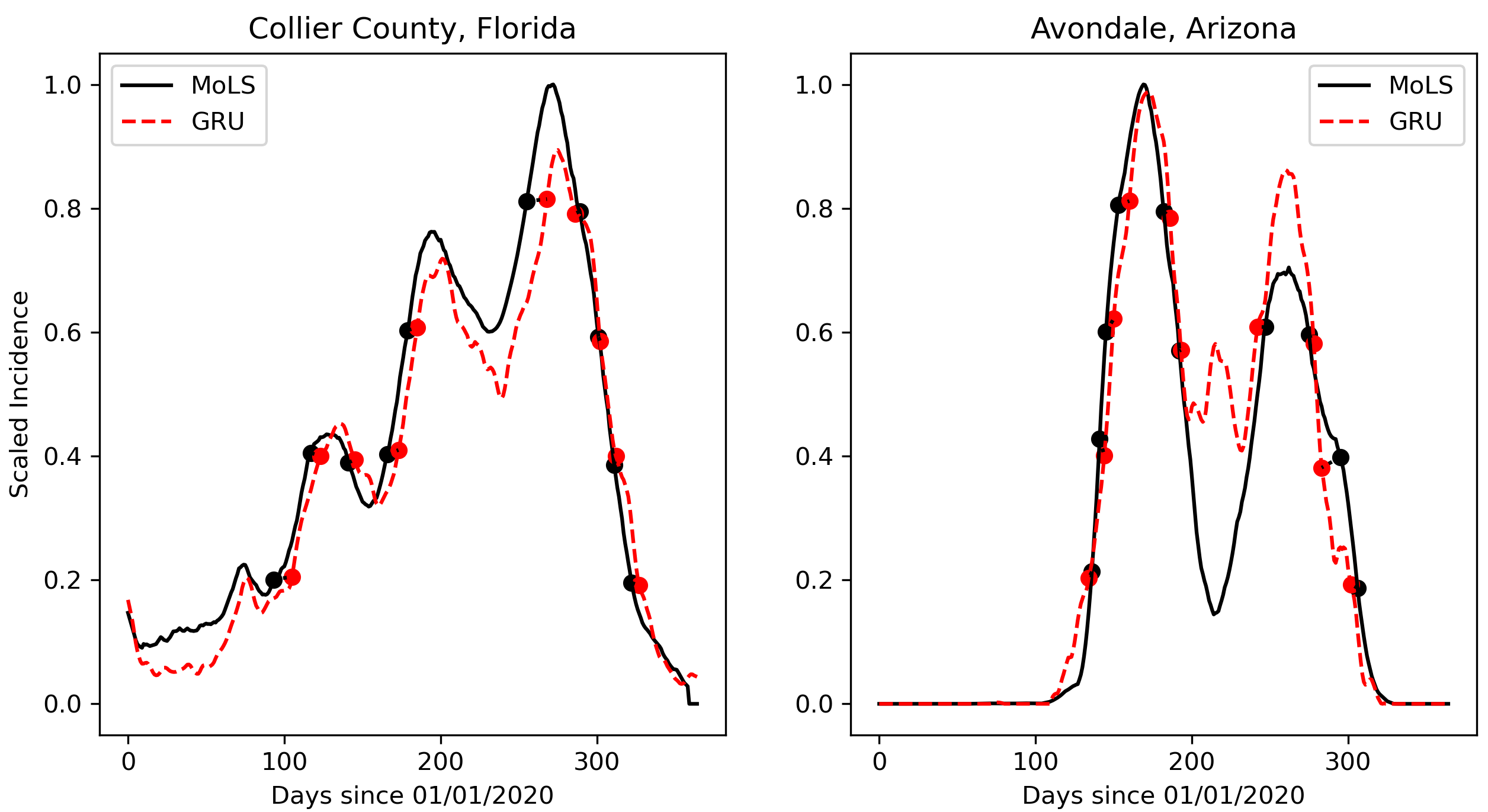}
    \caption{Observed (black solid curve) and predicted (red dashed curve for the GRU model) scaled abundance in Collier County, FL (\textit{left}) and Avondale, AZ (\textit{right}). To make it easier to visualize thresholds, each trace is scaled to the peak height of the observed (MoLS) abundance. The dots mark the times when each time series reached 20\%, 40\%, 60\%, and 80\% of the maximum MoLS abundance. Points in matching pairs are connected by dotted lines, whose projection on the horizontal axis has length $D_{on}$ or $D_{off}$. Black (resp. red) dots that are not matched to a red (resp. black) dot are omitted in this figure for clarity.}
    \label{fig:on-off}
\end{figure}

The season feature metrics quantify differences in the observed time-frames at which certain thresholds of mosquito abundance are reached for the target MoLS data and the ANN reconstructions. These are calculated in two steps: given a threshold $T$, we first identify a set of time intervals $(i_m, j_m)_{T}$ when MoLS estimates stably remain above $T$. Similarly, we identify intervals $(\hat{\imath}_n, \hat{\jmath}_n)_{T}$ when the ANN estimates remain above $T$. See Appendix \ref{sec:Feature Identification} for details. We then quantify the agreement between the two sets of intervals by calculating the differences $D_{on} \approx \hat{\imath}_n - i_m$ and $D_{off} \approx \hat{\jmath}_n - j_m$ in onset values (when MoLS and the ANN predictions first remain above $T$) and offset values (when predictions return below $T$). We use approximation symbols here because multiple intervals need to be properly matched to one another in order to calculate $D_{on}$ and $D_{off}$. Details are provided in Appendix \ref{sec:Feature Performance Identification}.

To assess the ability of each ANN to reproduce the results of MoLS, we choose to generate results for thresholds $T$ that are proportional to MoLS data; we test predictions for thresholds at 20\%, 40\%, 60\%, and 80\% of MoLS peak height to capture differences in both peak timing and season length. This process is illustrated in Figure \ref{fig:on-off} for the GRU model in two locations: Collier County, Florida and Avondale, Arizona. For a given year, the season length $\ell_S$ is defined as $\ell_S = \max(j_{m}) - \min(i_{m})$, where the $j_m$ and $i_m$ time points are calculated for $T = 20\%$ of the MoLS peak value. Onsets and offsets are then scaled by the average of $\ell_S$ over the testing years at the given location, to make errors relative to the environment in which they occur; if the location in consideration has a longer average season length, the relative seasonal difference will be lower than the same absolute seasonal difference for a location with a shorter average season length.
For each combination of model, threshold value, and testing location we report the means $\bar D$ and standard deviations $\sigma(D)$ of (the scaled) $D_{on}$ and $D_{off}$ values over all years. Additionally, we record the total number of times each neural network model did not reach the selected threshold.

Finally, we introduce a composite score, $S$, defined in Appendix \ref{sec:Combined Score}. This number, which takes into account performance on global and seasonal metrics, provides a convenient way of comparing models, and is used to rank the models in Figure \ref{fig:combined_score}. It penalizes errors in mean values as well as error variability. Low values of $S$ are associated with what the authors view as good overall ability for an ANN model to reproduce MoLS results. 

\section{Results}

\subsection{Performance of Base Models}\label{sec:Base Model Performance}

During the model development process we use a $MSE$ loss function to train the model parameters and $R^2$ accuracy to assess overall performance. As described in \S\ref{sec:Hyperparameter Selection}, we vary the hyperparameters, mimicking a grid search, until we observe diminishing returns on validation performance. The performance metrics for our trained models with the final selected hyperparameter values are shown in Table \ref{tab:verification} for each of the three training, validation, and testing subsets (defined in \S\ref{sec:Data Processing}). The output abundance curves are smoothed according to \S\ref{sec:Data Smoothing}, and the $R^2$ and $RMSE$ metrics for each subset are calculated on the entire output vectors. In particular, the training metrics reflect performance on the entire training subset, even though only 1000 input samples per location were used during the training process (\S\ref{sec:Data Processing}). Although the $RMSE$ values may seem high, the corresponding mean absolute errors are around half as large (data not shown). Thus, on the testing subset the GRU model differs from MoLS by an absolute average of $\approx57$ mosquitoes per day, a reasonable value when compared to MoLS abundance peak heights, which range from several hundreds in Arizona to several thousands in Florida (MoLS time series for a period of 9 years at the testing locations are provided as supplementary material). Later, we use the $NRMSE$ metric, which estimates error relative to local abundance values.

\begin{table}[ht]
    \centering
    \begin{tabular}{|c|c|c|c|c|}
         \hline
         \textbf{Model} & \textbf{Metric} & \textbf{Training} & \textbf{Validation} & \textbf{Testing}  \\

	\hline 
	\multirow{2}{*}{Baseline} & $RMSE$ & 481.118 & 466.243 & 508.648\\ 
	& $R^2$ & 0.558 & 0.551 & 0.558\\ 

	\hline 
	\multirow{2}{*}{FF} & $RMSE$ & 131.002 & 158.352 & 151.999\\ 
	& $R^2$ & 0.967 & 0.948 & 0.961\\ 

	\hline 
	\multirow{2}{*}{LSTM} & $RMSE$ & 75.783 & 135.0 & 131.088\\ 
	& $R^2$ & 0.989 & 0.962 & 0.971\\ 

	\hline 
	\multirow{2}{*}{GRU} & $RMSE$ & \textbf{62.788} & \textbf{131.073} & \textbf{116.376}\\ 
	& $R^2$ & \textbf{0.992} & \textbf{0.965} & \textbf{0.977}\\
	
         \hline
    \end{tabular}
    \caption{$RMSE$ and $R^2$ metrics for the training, validation, and testing data subsets. The best performing values of $RMSE$ and $R^2$ for each subset are in bold. }
    \label{tab:verification}
\end{table}

Table \ref{tab:verification} shows that the ANN models perform well at replicating the mosquito abundance predictions of MoLS, with the GRU model being the best performer overall. The gap between the training and validation performance indicates slight over-fitting to the training subset, but the results of training and validation are satisfactory, with $R^2$ values above 0.96 and 0.94 respectively. Further, $R^2$ values greater than 0.96 are achieved on the testing subset. The performance of the baseline model (first row of Table \ref{tab:verification}) is clearly sub-par. This is not surprising since MoLS predictions result from a complex process that is unlikely to be captured by a linear model.
In what follows, we provide a detailed analysis of the models performance on the testing subset, using the metrics defined in Section \ref{sec:New Metrics Section}. The results are expected to be representative of what a future user would experience, since they apply to data that were not used during the training and validation process. 

\begin{figure}[ht]
    \centering
    \includegraphics[width=\textwidth]{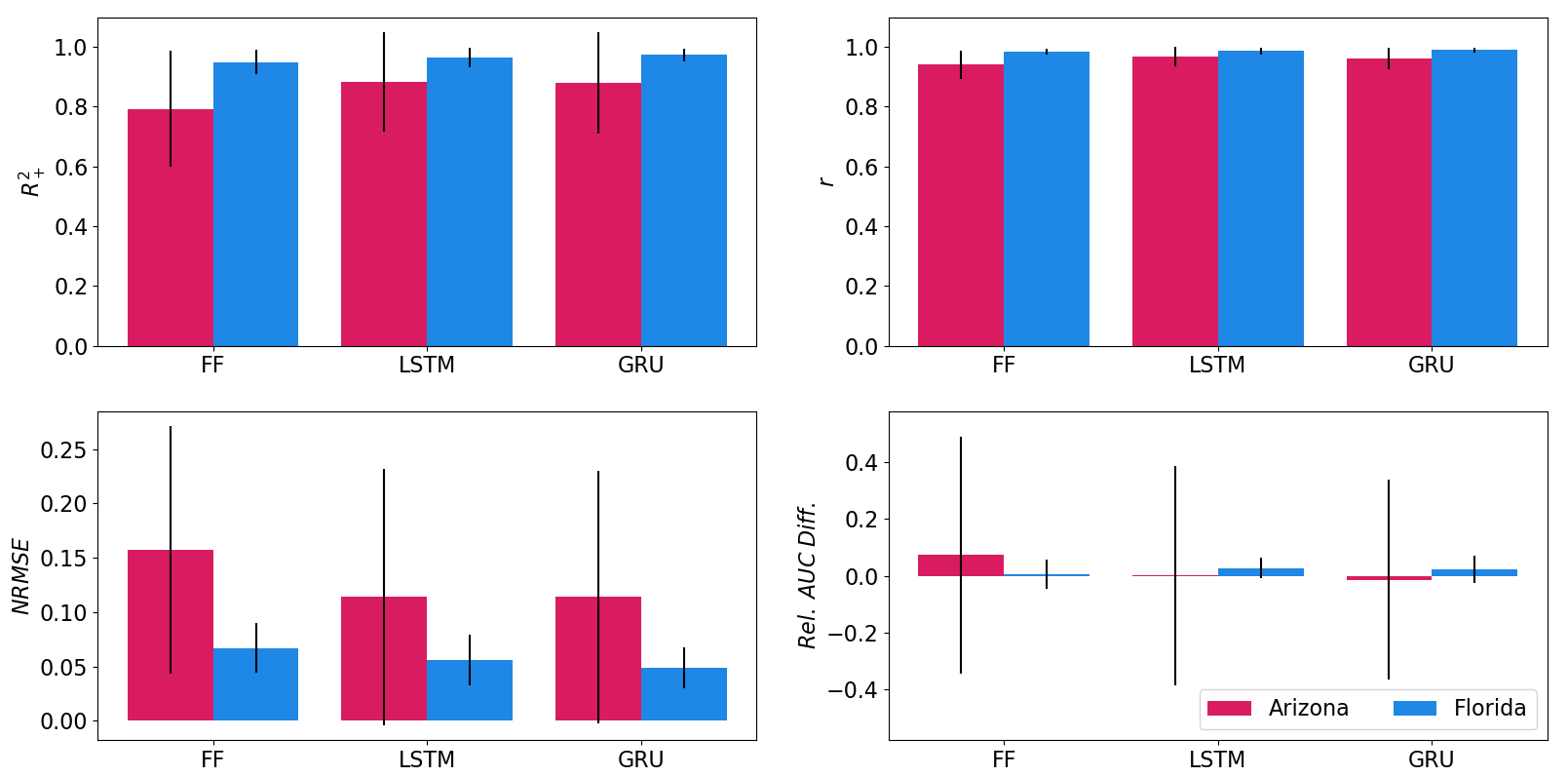}
    \caption{Global fit metrics for Arizona (red, left-hand columns) and Florida (blue, right-hand columns). The thin vertical lines have length equal to two sample standard deviations. In each state, all locations and years available in the testing subset were used. See \S\ref{sec:New Metrics Section} and Appendix \ref{sec:Global Perf Metrics} for a description of the metrics.}
    \label{fig:global_metrics_az_fl}
\end{figure}

Table \ref{tab:base_global} expands on Table \ref{tab:verification} and estimates the global fit metrics for the three models. The sample means reported in the $R_+^2$ column are lower than the $R^2$ score in Table \ref{tab:verification} because the present analysis is performed at a more granular level, for each location and year, rather than over the entire testing subset. The performance of ANN models is still quite good overall, but the relatively large standard deviations associated with the $Rel.\ AUC\ Diff$ metric indicate variability in the way their output fits MoLS results. Figure \ref{fig:global_metrics_az_fl} illustrates the nature of this variability by comparing model performance in two states with low (Arizona) and high (Florida) abundance numbers. The consistent scores in the latter (small standard deviations for all metrics) and differences between the two states (worse performing mean values associated with larger standard deviations in Arizona) suggest that temporal-variability plays less of a role here than location-variability. 

\begin{table}[ht]
\centering
\small
\begin{tabular}{|c|c|c|c|c|c|}
\hline
\multirow{2}{*}{\textbf{Model}} & \multicolumn{4}{c|}{\textbf{Metric}} \\
\cline{2-5}
& $R_+^2 \uparrow$ & $NRMSE \downarrow$ & $Rel.\; AUC\; Diff. \downarrow$ & $r\uparrow$ \\
	\hline 
	Baseline & 0.301 $\pm$ 0.342 & 0.381 $\pm$ 0.27 & -1.077 $\pm$ 1.294 & 0.737 $\pm$ 0.296\\ 

	\hline 
	FF & 0.871 $\pm$ 0.142 & 0.112 $\pm$ 0.073 & 0.09 $\pm$ 0.239 & 0.961 $\pm$ 0.038\\ 

	\hline 
	LSTM & 0.916 $\pm$ 0.118 & 0.088 $\pm$ 0.07 & \textbf{0.031} $\pm$ \textbf{0.224} & 0.974 $\pm$ 0.028\\ 

	\hline 
	GRU & \textbf{0.923} $\pm$ \textbf{0.119} & \textbf{0.083} $\pm$ \textbf{0.071} & 0.036 $\pm$ 0.207 & \textbf{0.975} $\pm$ \textbf{0.029}\\ 

\hline
\end{tabular}
\caption{Global fit metrics calculated for the testing subset. The arrows point in the direction of more desirable magnitudes. The entries for metric $*$ are formatted as $\bar{*}\pm\sigma(*)$ where $\bar{*}$ and $\sigma(*)$ are the mean and standard deviation calculated over all locations and years. In each column, the entry in bold has the best performing mean. See \S\ref{sec:New Metrics Section} and Appendix \ref{sec:Global Perf Metrics} for a description of the metrics.}
\label{tab:base_global}
\end{table}

\begin{table}[h!]
\small
\centering
\begin{tabular}{|c|c|c|c|c|c|}
     \hline
     \multicolumn{6}{|c|}{\textbf{Arizona}}\\
     \hline
     \multirow{2}{*}{\textbf{Model}} &  \multirow{2}{*}{\textbf{Metric}} & \multicolumn{4}{c|}{\textbf{Threshold (\% of Max MoLS Prediction)}} \\
     \cline{3-6}
     & & \textbf{20\%} & \textbf{40\%} & \textbf{60\%} & \textbf{80\%} \\
     \hline 
    \multirow{2}{*}{FF} & $D_{on}$ & -0.011 $\pm$ 0.051 & -0.021 $\pm$ 0.055 & -0.055 $\pm$     0.086 & -0.028 $\pm$ 0.206\\ 
    & $D_{off}$ & \textbf{0.0} $\pm$ \textbf{0.066} & 0.015 $\pm$ 0.078 & 0.06 $\pm$ 0.126 & 0.092 $\pm$ 0.2\\

    \hline 
    \multirow{2}{*}{LSTM} & $D_{on}$ & \textbf{-0.004} $\pm$ \textbf{0.056} & \textbf{-0.006} $\pm$ \textbf{0.052} & \textbf{-0.03} $\pm$ \textbf{0.072} & \textbf{-0.043} $\pm$ \textbf{0.09}\\ 
    & $D_{off}$ & -0.005 $\pm$ 0.036 & \textbf{0.0} $\pm$ \textbf{0.094} & \textbf{0.01} $\pm$ \textbf{0.115} & 0.043 $\pm$ 0.108\\ 

    \hline 
    \multirow{2}{*}{GRU} & $D_{on}$ & -0.011 $\pm$ 0.049 & -0.013 $\pm$ 0.051 & -0.038 $\pm$ 0.111 & -0.032 $\pm$ 0.138\\ 
    & $D_{off}$ & -0.008 $\pm$ 0.047 & \textbf{0.0} $\pm$ \textbf{0.087} & 0.009 $\pm$ 0.134 & \textbf{0.012} $\pm$ \textbf{0.141}\\

     \hline
     \multicolumn{6}{|c|}{\textbf{Florida}}\\
     \hline
     \multirow{2}{*}{\textbf{Model}} &  \multirow{2}{*}{\textbf{Metric}} & \multicolumn{4}{c|}{\textbf{Threshold (\% of Max MoLS Prediction)}} \\
     \cline{3-6}
     & & \textbf{20\%} & \textbf{40\%} & \textbf{60\%} & \textbf{80\%} \\
    \hline 
    \multirow{2}{*}{FF} & $D_{on}$ & -0.019 $\pm$ 0.04 & -0.008 $\pm$ 0.032 & \textbf{0.0} $\pm$ \textbf{0.031} & \textbf{0.001} $\pm$ \textbf{0.06}\\ 
    & $D_{off}$ & 0.009 $\pm$ 0.029 & 0.005 $\pm$ 0.039 & -0.007 $\pm$ 0.048 & -0.032 $\pm$ 0.052\\ 

    \hline 
    \multirow{2}{*}{LSTM} & $D_{on}$ & \textbf{-0.013} $\pm$ \textbf{0.037} & -0.014 $\pm$ 0.021 & -0.023 $\pm$ 0.031 & -0.029 $\pm$ 0.05\\ 
    & $D_{off}$ & -0.003 $\pm$ 0.022 & \textbf{-0.003} $\pm$ \textbf{0.026} & -0.004 $\pm$ 0.023 & -0.01 $\pm$ 0.046\\ 

    \hline 
    \multirow{2}{*}{GRU} & $D_{on}$ & -0.017 $\pm$ 0.039 & \textbf{-0.012} $\pm$ \textbf{0.02} & -0.011 $\pm$ 0.024 & -0.02 $\pm$ 0.036\\ 
    & $D_{off}$ & \textbf{-0.001} $\pm$ \textbf{0.018} & -0.006 $\pm$ 0.033 & \textbf{-0.003} $\pm$ \textbf{0.016} & \textbf{-0.008} $\pm$ \textbf{0.038}\\ 

     \hline

\end{tabular}
\caption{Season feature metrics calculated for Arizona and Florida testing locations. Seasonal differences for a location and year are scaled by the average length of the season at the 20\% threshold. The entries, formatted as $\bar{D}\pm\sigma(D)$, are calculated over all locations and years in each state. Bold entries correspond to the lowest values of $\vert \bar D \vert \cdot \sigma(D)$ for each threshold, with $D = D_{on}$ or $D_{off}$. See \S\ref{sec:New Metrics Section} and Appendix \ref{sec:Feature Identification} for -a description of the metrics.}
\label{tab:az_fl_season_metrics}
\end{table}

We now turn to the season fit metrics, which explicitly capture deviations in the timing of the predicted season onset, offset, and peaks (see Figure \ref{fig:on-off}). Table \ref{tab:az_fl_season_metrics} shows these metrics for Arizona and Florida; as in Figure \ref{fig:global_metrics_az_fl}, averages are taken over all locations and years in each state. Although the sample means of $D_{on}$ and $D_{off}$ are typically low, they are better in Florida (less than a few percents of the season length) than Arizona (up to 9\% of the season length). Moreover, the standard deviations for Arizona are again larger than for Florida, especially at the 60\% and 80\% thresholds. This suggests that ANNs have more difficulties capturing the timing of peaks than season onsets and offsets (corresponding to the 20\% threshold). A similar trend is observed in the first three rows (above the double line) of Table \ref{tab:seasonal_perf_all}, Appendix \ref{app:B}, which shows the means and standard deviations of $D_{on}$ and $D_{off}$ over all locations and years in the testing subset.

All of the ANN models are trained on samples randomly selected from the training subset. As a consequence, different realizations of the same model, trained with the same hyperparameters but with different samples, will produce slightly different results. To illustrate this variability, Appendix \ref{App:other_models} provides performance results for the three base models trained either on a different, or on a larger, set of samples. The stable performance, as documented by the various tables presented in this appendix, suggests that the default hyperparameter values chosen during model development are appropriate.

The above analysis reveals that lower model performance may be associated with a lack of ability to capture the timing of abundance peaks. The next section explores whether training the models on augmented datasets that specifically address how they respond to high and low temperatures improves consistency.

\subsection{Performance of Variant Models}

\begin{figure}[ht]
    \centering
    \includegraphics[width=\textwidth]{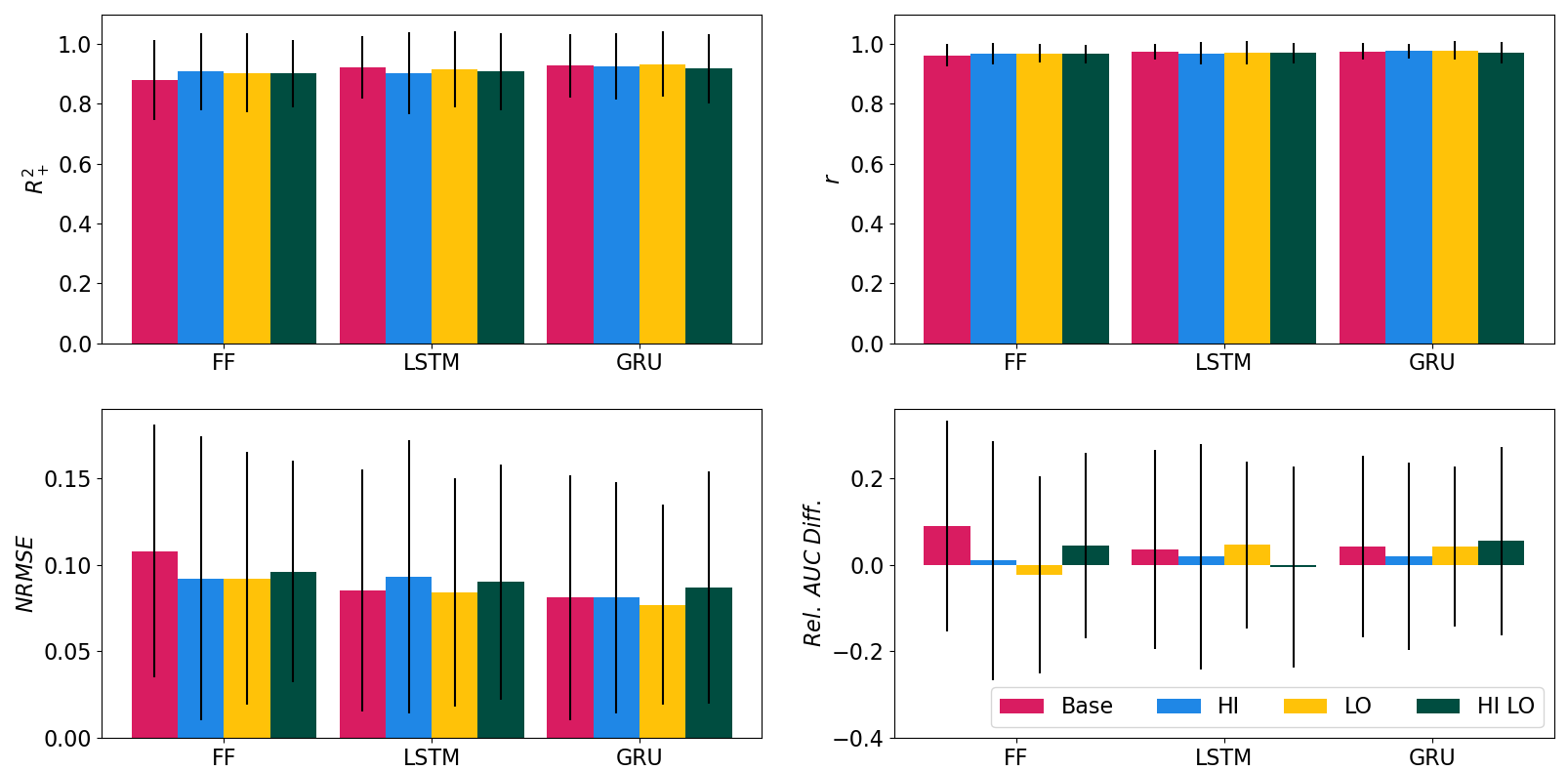}
    \caption{Average performance metrics and standard deviation of the testing locations. The legend is in the bottom right panel.}
    \label{fig:all_mean_metrics}
\end{figure}

The augmentation methods only lead to nominal improvement in the global fit metrics, except possibly for lower $Rel.\ AUC\ Diff.$ values associated with the HI versions of each model. This is illustrated in Figure \ref{fig:all_mean_metrics}, which shows these metrics, calculated for all locations and years in the testing subset, for all models. In addition, the results of the seasonal feature analysis shown in Appendix \ref{app:B} (Table \ref{tab:seasonal_perf_all}) indicate slight performance improvement for the GRU variants, in particular at high threshold values.

Appendix \ref{app:C} contains a case study of the GRU variants for Avondale, Arizona and Collier County, Florida. We show the 2020 abundance curves, as well as associated global fit and seasonal feature metrics. The case study exemplifies differences in model variants and performance between the two locations. In particular, the HI variant is able to capture the dip in abundance due to hot summer temperatures. It should be noted however that variations in the abundance curves produced by the ANNs are minimal when compared to changes in MoLS dynamics due to weather data (see Figure \ref{fig:MoLS_AZ}) or even location (Figure \ref{fig:MoLS_CA}).

\subsection{Comparative Model Performance}\label{sec:Combined Results}

\begin{figure}[ht]
    \centering
    \includegraphics[width=\textwidth]{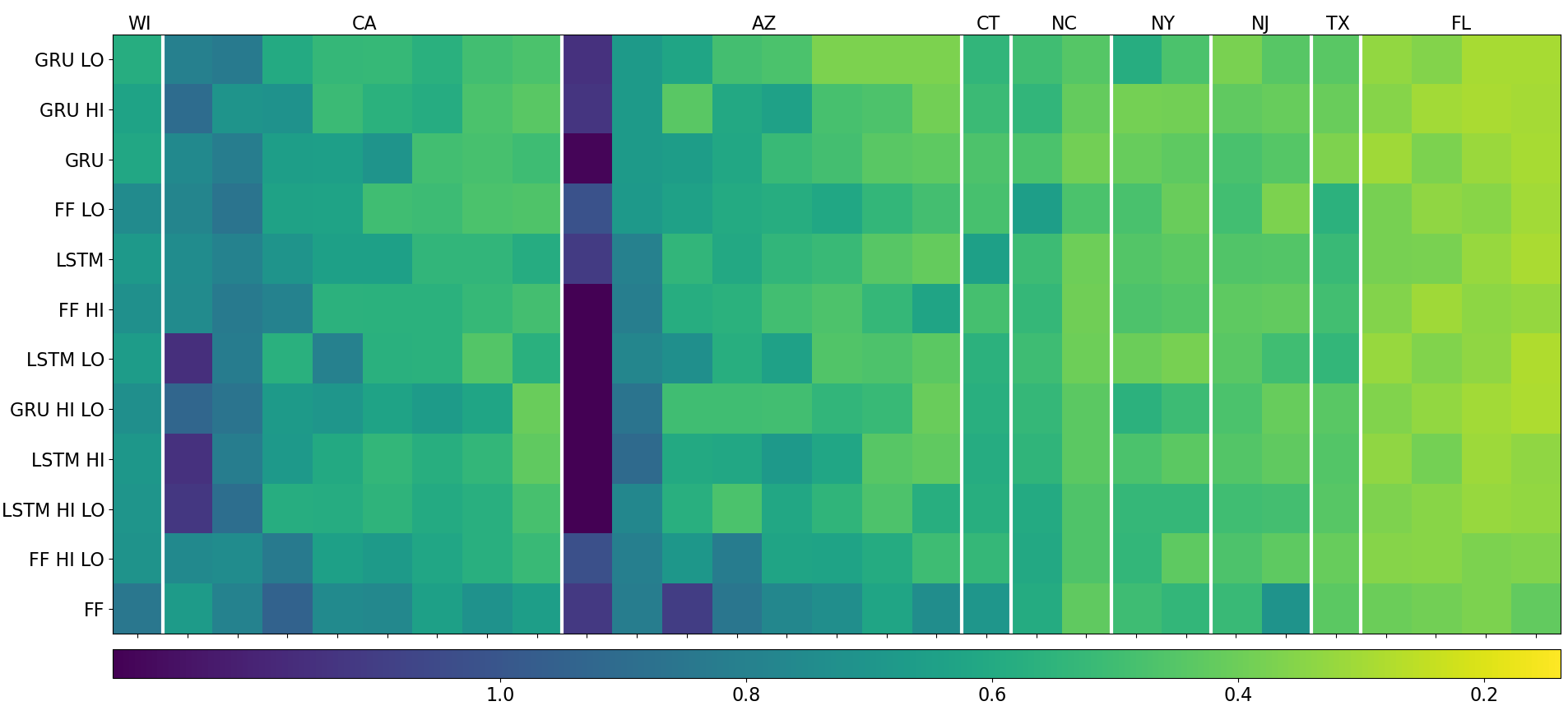}
    \caption{Combined scores on testing locations (see \S\ref{sec:Combined Score} for metric definition). The rows represent the models and are organized bottom to top from highest score (worst model) to lowest score (best model). The columns represent individual locations (see Table \ref{tab:testing_locs} for the names), and the vertical, white lines separate states. The states are organized left to right from highest mean score to lowest, and within each state the locations are organized left to right by descending score.}
    \label{fig:combined_score}
\end{figure}

The overall metric defined in Appendix \ref{sec:Combined Score} combines the global fit and season feature metrics into a single score. It provides a balanced picture of the performance of each model by taking into account accuracy in terms of season abundance, season length, and peak timing. We show these results in Figure \ref{fig:combined_score}. The GRU, which was the best of the three base models, is only outperformed by its HI and LO variants. Similarly, the FF model is outperformed by all of its variants. On the other hand, the LSTM variants led to a loss of performance, as well as did most of the HI LO variants.
Also evident from this figure is the significant impact of location on model performance. In particular, all models score poorly for the left-most location in Arizona (Fortuna Foothills). Further inspection of the associated time series reveals a significant decrease in MoLS abundance starting at the end of 2018 which is not matched by the ANNs. We believe this is due to a large spike in precipitation (more than 400 times the average daily value) included in the MACA data set on 10/24/2017, which is then followed by months of lower-than-average rainfall. MoLS takes into account the possibility that rain creates new habitat (pools of water) where mosquitoes can develop. These come from a reserve of eggs that are available in the environment and can hatch in newly created breeding grounds. In the case of excessive rainfall, especially in regions of low mosquito abundance, the pool of eggs may be exhausted in a single event. A drop in future abundance can ensue since not all recently hatched eggs will survive to adulthood. In addition, during periods of dry weather, a depleted reserve of eggs is likely to take a long time to rebuild to normal levels. If large precipitation events occur outside of the $\Delta$ days window used for input, the ANNs cannot be aware of them and, as a consequence, will produce results reflecting normal abundance given the local weather conditions. For reference, the time series for MoLS and the GRU HI model at the locations listed in Figure \ref{fig:combined_score} are provided as supplementary material. In the case of Fortuna Foothills, the ANN correctly reproduces the double-peak pattern seen in MoLS results, but has higher abundance in 2019-2020. This is consistent with our proposed explanation that the conjunction of a rare high precipitation event and overall low pre-event abundance levels (less than 1000 mosquitoes at peak height) led to a crash in MoLS population estimates that is not captured by the ANNs.

Appendix \ref{app:CC} presents an analysis of the performance of the GRU HI model on the Capital Cities dataset, which covers 9 years of data over 44 locations. The map of Figure \ref{fig:capitals_score_map} indicates that the model works very well in the eastern and southeastern regions of the United States, but has inferior performance in the west. Arizona and California are not included because the corresponding locations (Phoenix and Sacramento) are in the principal dataset (see Figure \ref{fig:combined_score} for results). In addition, we observe a strong correlation between abundance and performance: regions of low or irregular MoLS abundance are typically associated with worse ANN performance. This is illustrated in the top row of Figure \ref{fig:GRUHI_MoLS_Comparison}, for Nevada and Montana, which have abundance peak heights of only a few hundred mosquitoes.

\section{Discussion}

MoLS, a stochastic agent-based model for {\it Aedes aegypti} abundance that was validated against surveillance data in Puerto Rico \cite{lega2017aedes}, uses weather data to simulate the life cycle of a large number of mosquitoes and estimate expected daily abundance. It is natural to ask whether a properly trained artificial neural network is able to ``learn'' how MoLS combines weather-dependent development, survival, and reproductive rates to make its predictions. In this paper we demonstrate that it is possible to train ANNs that map meteorological data to mosquito numbers in a way consistent with MoLS results. Although the 12 models considered here achieve varying levels of success, they are generally able to replicate the trends observed in MoLS time series, indicating a neural network can function as an equation-free model of {\it Aedes aegypti} abundance. 

As shown by the sub-par performance of the baseline model, learning how MoLS functions requires a more complex setup than a linear regression. While all three base ANN models use the same architecture, shown in Figure \ref{fig:arch}, incorporating recurrent layers (LSTM and GRU layers) improves performance, compared to the FF model (Table \ref{tab:verification}). This suggests that a model using the spatial feature extraction of convolution layers alone is unable to fully identify the relationship between weather features and mosquito abundance, and that combining the sequential ``memory'' feature of recurrent layers with the convolution layers better captures this relationship.

The metrics shown in Tables \ref{tab:base_global} and \ref{tab:seasonal_perf_all} indicate all ANN models have high overall skill, with minimal differences in global level performance between them. At a more granular level, the case study of Appendix \ref{app:C} suggests the GRU HI model is better able to capture abundance in hot summer months. Such an improvement is expected to be reflected in the composite metric of Appendix \ref{sec:Combined Score}, which by design is sensitive to variability in local performance. Indeed, Figure \ref{fig:combined_score} indicates the HI and LO data augmentation methods improve the performance of the GRU and FF models, although not that of the LSTM model. It is not clear at this point why the HI LO models are inferior to their HI and LO counterparts. The most probable explanation is that the data augmentation puts too much emphasis on extremes compared to typical temperatures, thereby lowering performance in generic situations. 

The performance of the GRU HI model on the Capital Cities dataset (Appendix \ref{app:CC}) reveals that ANNs perform well in regions where mosquito abundance is high - which is principally where they would be expected to be used. Each year of abundance predictions presented in this paper took an ANN only 0.33 seconds to generate using a laptop with a 1.6 GHz Dual-Core Intel Core i5 processor with 8GB RAM (compare to the 10 mn needed by MoLS to generate 10 years of abundance estimates on a single HPC core) and approximately 1-2 minutes to train per epoch for each model using a laptop with Intel Core i7-9750H CPUs with 16GB RAM and a single Nvidia GeForce GTX 1650 GPU with 4GB memory, depending on the sampling size and model architecture. The combination of speed and accuracy demonstrated in this article therefore identifies neural network models as top contenders for efficiently converting weather data into {\it Aedes aegypti} and more generally mosquito abundance. To encourage such applications of ANNs, all of our code is freely available on GitHub to researchers interested in improving on the present results. However, before using an ANN as a replacement for MoLS, its performance should be assessed with metrics similar to those presented in this article and the model that best captures local circumstances (e.g. the effect of hot summers on mosquito populations in Arizona) should be selected. A quantitative comparison with actual surveillance data is also recommended. Both MoLS and its ANN replacements introduced here produce daily numbers of scaled mosquito abundance. As previously mentioned, ``scaled'' means that the estimates are up to a multiplicative factor that depends on location, but not on time. If surveillance data are available, the value of this factor can be found via linear regression of MoLS results against the data, as was done in \cite{lega2017aedes}. The resulting estimate will depend on the type of mosquito traps used for surveillance, as well as on local considerations, such as the number of available breeding sites. Once rescaled, MoLS or ANN predictions should be able to capture overall abundance trends fairly well.

Possible applications include the use of local weather data for vector control interventions. In this case, abundance trends could be estimated on a daily basis and supplement routine surveillance; reliable weather data, as well as the best performing ANNs should be selected, to increase confidence in the results. Model limitations should also be taken into account. As suggested by the Fortuna Foothills example, extreme weather events that are localized in time but affect average population levels in the long term are not taken into account by the ANNs when such events fall outside the range of their input window of $\Delta$ days. However, if the user knows that such an event occurred, it is not difficult to recalibrate the ANNs by recalculating the scaling factor that relates their output to local surveillance data. More qualitative, longer-term planning, based on climate scenarios, should also be possible with the ANNs presented here, since estimating general, weather-based trends of mosquito abundance would suffice in that case. In addition, because \textit{Aedes aegypti} is a known vector for diseases like dengue, chikungunya, and Zika, many studies have provided environment suitability maps for this species and have used them to estimate disease risk (see for instance \cite{kraemer2015global, Messina16} and references therein). An ANN trained on reproducing MoLS predictions would make it possible to create similar maps from a weather-based mechanistic abundance model without the need of high power computing (HPC)
typically required to generate the same quantity of predictions using MoLS. Finally, the speed afforded by ANNs could allow the creation of interactive web apps able to produce estimates of mosquito abundance from local weather data at a user's request.

Another important application of abundance models is the development of probabilistic forecasts. Going beyond the point estimates produced by the models discussed in this article requires additional uncertainty quantification, especially in terms of the variability inherent to the local weather forecast used to make predictions (an example of how different models affect MoLS output is provided in Appendix \ref{app:A1}). Looking forward, we believe that when combined with assimilation of weather and surveillance data, the ANNs trained for this article can effectively contribute to the development of probabilistic mosquito abundance forecasting models. We leave this for future work.

\section*{Acknowledgements}

\subsubsection*{Author Contributions}
JL conceived of the project and provided the MoLS data. SC led model development efforts and AK led the performance analysis of the models. All authors contributed to the analysis of the results and to the writing of the manuscript. All authors approved the final version of this article.

\subsubsection*{Funding \& Research Support}
Research reported in this publication was supported in part by the National Institute of General Medical Sciences of the National Institutes of Health under grant GM132008 and by the University of Arizona’s BIO5 Institute Team Scholars Program. An allocation of computer time from the UA Research Computing High Performance Computing (HPC) at the University of Arizona is gratefully acknowledged.

\subsubsection*{Competing Interests}
All authors declare that they have no competing interests.

\subsubsection*{Data and Code Availability}

All data files, trained models, and codes used to create the figures are available at \url{https://github.com/T-MInDS/Aedes-AI}

\newpage

\appendix
\section{Weather and MoLS Time Series}
\label{app:A1}
\begin{figure}[h]
    \centering
    \includegraphics[width=.8\textwidth]{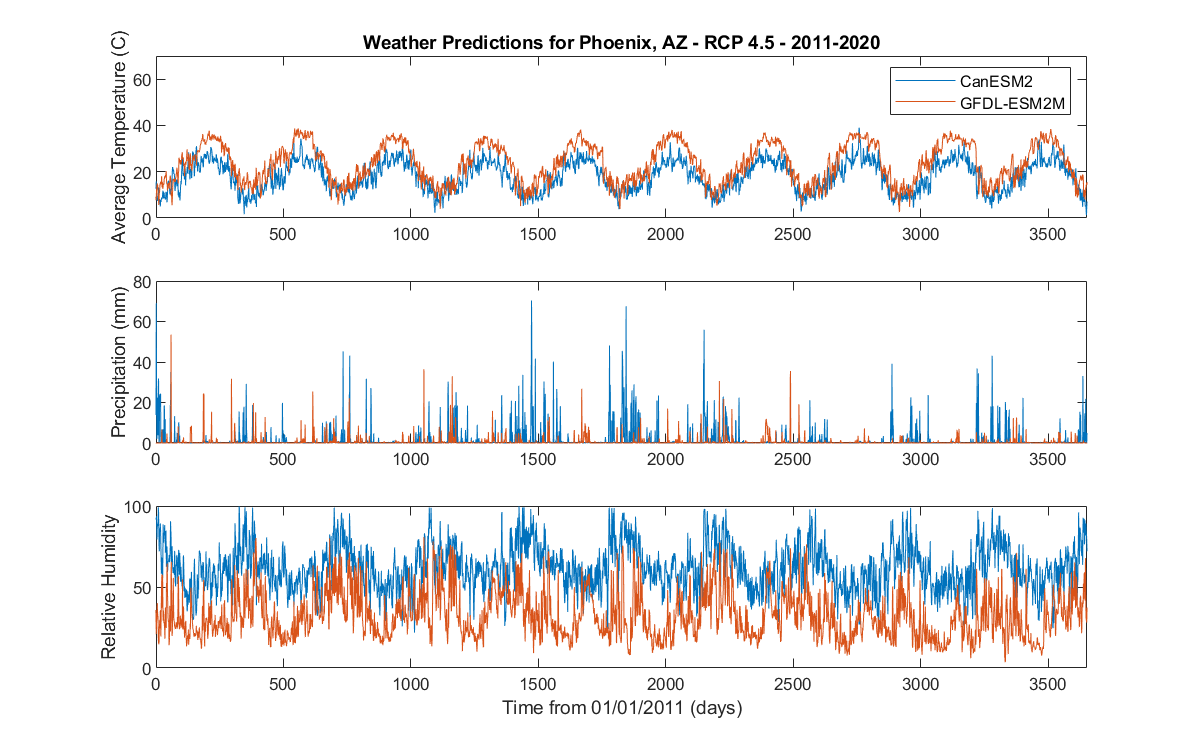}
    \caption{\label{fig:Weather_AZ} MACA weather time series from 01/01/2011 to 12/31/2020 for Phoenix, AZ, based on two different climate models: GFDL-ESM2M and CanESM2. Top: average daily temperature; Middle: daily precipitation in millimeters; Bottom: relative humidity.}
\end{figure}

Figure \ref{fig:Weather_AZ} illustrates daily fluctuations and seasonal patterns observed in typical input weather data. MACA estimates of average daily temperature, precipitation, and relative humidity are plotted for two different climate models, GFDL-ESM2M \cite{US_Model} and CanESM2 \cite{Can_Model}.

\begin{figure}[h]
    \centering
    \includegraphics[width=.8\textwidth]{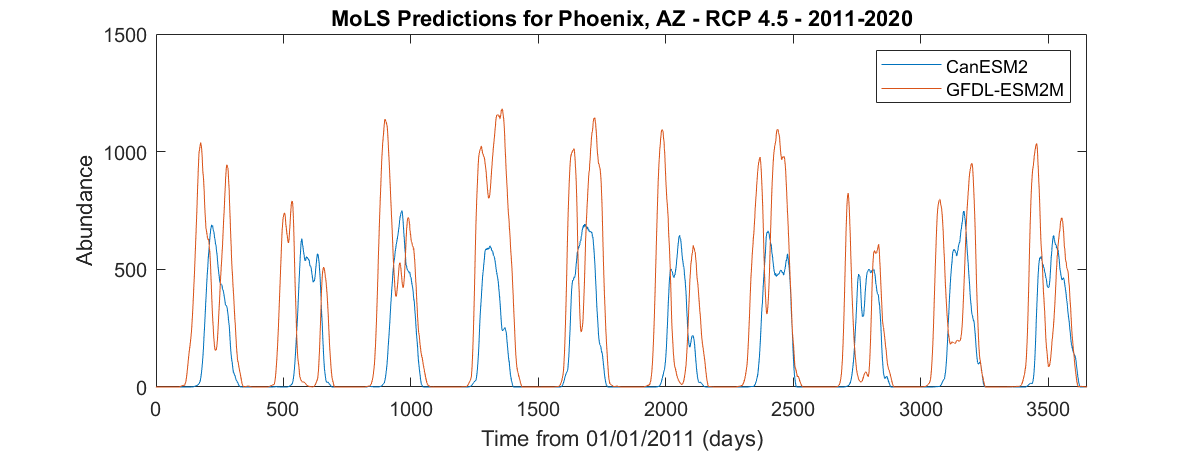}
    \caption{\label{fig:MoLS_AZ} Time series of gravid female abundance generated by MoLS using the weather data shown in Figure \ref{fig:Weather_AZ}.}
\end{figure}

Visible differences between the two models lead to differences in MoLS predictions, as illustrated in Figure \ref{fig:MoLS_AZ}: hotter and less humid conditions in the summer lead to double peaks in estimated mosquito abundance and to longer mosquito seasons.

\begin{figure}[h]
    \centering
    \includegraphics[width=.8\textwidth]{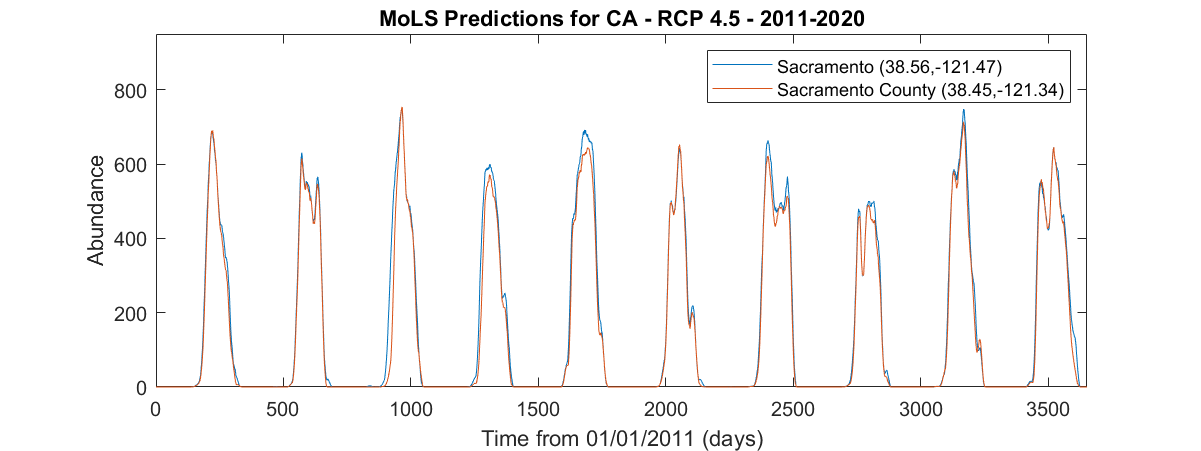}
    \caption{\label{fig:MoLS_CA} Time series of gravid female abundance generated by MoLS for two different locations near Sacramento, CA.}
\end{figure}

MoLS output is sensitive to changes in location due to changes in associated weather data. Figure \ref{fig:MoLS_CA} shows MoLS predictions for Sacramento, CA (latitude: 38.56, longitude: -121.47) and for the centroid of Sacramento County (latitude: 38.35, longitude: -121.34). Note the differences in peak height (years 4, 5, 7, and 9) and in season length (years 1, 3, 4, and 10).

\begin{figure}[ht]
    \centering
    \includegraphics[width=0.9\textwidth]{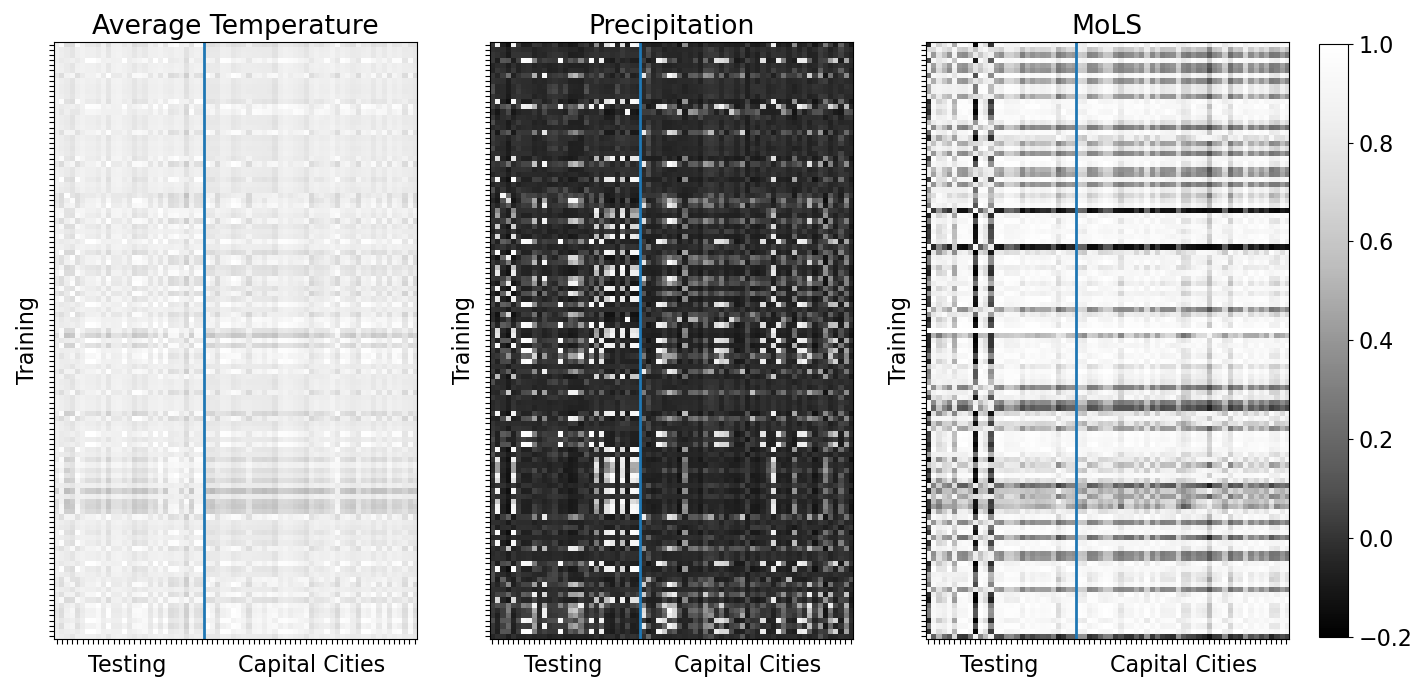}
    \caption{Pearson correlation between training and testing data for average temperature, precipitation, and MoLS estimates in calendar year 2016. The correlations between training and testing locations are shown to the left of the blue line, and the correlations between training locations and capital cities are shown to the right. Each tick mark represents one location in the subset.}    \label{fig:location_correlations}
\end{figure}

Figure \ref{fig:location_correlations} shows the correlation between the 2016 training and testing locations (\textit{left} of the blue line), and 2016 training locations and capital cities (\textit{right} of the blue line) for average temperature, precipitation, and MoLS predictions. The average temperature is seasonal, and thus, highly correlated among all locations, regardless of the relative temperature scale between locations. The low correlation for precipitation indicates that the ANN models are tested on samples with different weather features than the training samples. These differences in weather trends lead to moderately correlated annual MoLS predictions.

\clearpage

\newpage

\section{Description of ANN Layers}
\label{app:A}
This appendix provides additional details for the layers used in the neural network models described in \S\ref{sec:Models}. Note that the parameters appearing in the weight matrices and kernels are learned during the training phase, as described in \S\ref{sec:Model Training}. Once trained, the models and their saved parameters may be used for any viable input sample. For a more thorough description of neural networks, we direct readers to \cite{Goodfellow-et-al-2016}. Table \ref{tab:vars} shows the variables used in the convolutional layers (ConvLayers), the LSTM layers, and the GRU layers. Each layer is described in more details below.

\begin{table}
\centering
\begin{tabular}{|c|c|c|c|}
\hline
     \textbf{Variable} & \textbf{Description} & \textbf{Dimension/Value} & \textbf{Layers using Variable}  \\
     \hline
     $l$ & Layer & --- & All \\
     $\boldsymbol{x}_l$ & Input for layer $l$ & $[n_l \times m_l]$ & All \\
     $n_f$ & Number of filters & $64$ & ConvLayers \\
     $f$ & Filter & $f\in[1,n_f]$ & ConvLayers \\
     $\boldsymbol{K}_f$ & Kernel for filter $f$ & $[3\times m_l]$ & ConvLayers \\
     $t$ & Time index of $x_l$ & $t\in[0,n_l)$ & All \\
     ${\boldsymbol O}^C_l$ & Output matrix for ConvLayer $l$ & $[(n_l-2) \times 64]$ & ConvLayers \\
     $n_u$ & Length of hidden state & $64$ & LSTM/GRU Layers \\
     $\boldsymbol{c}_t$ & Cell state at time $t$ & $[1\times n_u]$ & LSTM Layers \\
     $\boldsymbol{h}_t$ & Hidden state at time $t$ & $[1\times n_u]$ & LSTM/GRU Layers \\
     $\boldsymbol{f}_t$ & Forget gate at time $t$ & $[1\times n_u]$ & LSTM Layers \\
     $\boldsymbol{i}_t$ & Input gate at time $t$ & $[1\times n_u]$ & LSTM Layers \\
     $\boldsymbol{o}_t$ & Output gate at time $t$ & $[1\times n_u]$ & LSTM Layers \\
     $\boldsymbol{W}_*$ & Weight matrix for gate $*$ & $[(m_l+n_u) \times n_u]$ & LSTM/GRU Layers \\
     $\boldsymbol{b}_*$ & Bias for gate $*$ & $[1\times n_u]$ & LSTM/GRU Layers \\
     $\boldsymbol{r}_t$ & Reset gate at time $t$ & $[1\times n_u]$ & GRU Layers \\
     $\boldsymbol{z}_t$ & Update gate at time $t$ & $[1\times n_u]$ & GRU Layers \\
     \hline
\end{tabular}
\caption{Variables used in the convolution layers, LSTM layers, and GRU layers. ``All" means the variables were used in the three layer types.}
\label{tab:vars}
\end{table}

\subsection{Convolution Layers}

Let $\boldsymbol{x}_l \in \mathbb{R}^{n_l\times m_l}$ be the input for layer $l$ and $\boldsymbol{K}_f \in \mathbb{R}^{n_K \times m_l}$ be the kernel for filter $f\in[1,n_f]$. The dimensions of $\boldsymbol{x}_l$ are dependent on layer $l$, but the dimensions of $\boldsymbol{K}_f$ are constant for all filters $f$. In this article, $n_K = 3$. Let $t$ be the time index of the input data such that $t\in[0,n_l)$. The output matrix ${\boldsymbol O}_l^C \in {\mathbb R}^{(n_l-n_K+1)\times n_f}$ of convolution layer $l$ is constructed as follows. For $t=0,...,(n_l-n_K)$ and $f=1,...,n_f$, let $\boldsymbol{\hat{x}}=\boldsymbol{x}_l[t:t+(n_K-1),:] \in \mathbb{R}^{n_K \times m_l}$ (note that the $t$ and $t+(n_K - 1)$ indices are inclusive). Then, ${\boldsymbol O}_l^C[t,f]=\sum_{a,b} [\boldsymbol{K}_f * \boldsymbol{\hat{x}}]_{a,b}$, where $*$ is component-wise multiplication, $a$ ranges from 0 to $n_K-1$, and $b$ from 0 to $m_l-1$. Table \ref{tab:FF dims} shows the input and output dimensions for the convolution layers in the models discussed in \S\ref{sec:Models}.

    \begin{table}[ht]
    \centering
    \begin{tabular}{c|l}
         \textbf{Layer} & \textbf{Input $\to$ Output Dimensions} \\
         \hline
         ConvLayer 1 & $[90\times 4] \to [88 \times 64]$ \\
         \hline
         ConvLayer 2 & $[88 \times 64] \to [86 \times 64]$ 
    \end{tabular}
    \caption{Input/output dimensions of the convolution layers used in the models described in \S\ref{sec:Models}.}
    \label{tab:FF dims}
    \end{table}
    
\subsection{LSTM Layers}
    
    The LSTM layers contain an LSTM unit with hidden state length $n_u=64$ and have two sources of information flow: $\boldsymbol{C}_t \in \mathbb{R}^{1\times n_u}$, the cell state, which intuitively may be thought of as the long-term memory, and $\boldsymbol{h}_t \in \mathbb{R}^{1\times n_u}$, the hidden state, which is a filtered version of the cell state. Here $t\in[0,n_l)$, and $\boldsymbol{x}_l \in \mathbb{R}^{n_l\times m_l}$ is the input for the LSTM layer $l$. Two of the gates in the unit, the forget gate ($\boldsymbol{f}_t \in \mathbb{R}^{1\times n_u}$) and the input gate ($\boldsymbol{i}_t \in \mathbb{R}^{1\times n_u}$), selectively update the cell state. The third gate, the output gate ($\boldsymbol{o}_t\in \mathbb{R}^{1\times n_u}$), filters the cell state to produce the output, hidden state $\boldsymbol{h}_t$.
    
    The above matrices are constructed as follows. First, the information from the previous hidden state, $\boldsymbol{h}_{t-1}$, is concatenated with the input at time $t$, $\boldsymbol{x}_t \in \mathbb{R}^{1 \times m_l}$, to obtain $(\boldsymbol{x}_t\| \boldsymbol{h}_{t-1}) \in \mathbb{R}^{1\times (m_l + n_u)}$, where $(\cdot\|\cdot)$ is the concatenation operator. Next, the gates select the information from the concatenation $(\boldsymbol{x}_t\| \boldsymbol{h}_{t-1})$ used to update the cell and hidden states. Specifically, $\boldsymbol{f}_t=\sigma((\boldsymbol{x}_t\| \boldsymbol{h}_{t-1})\boldsymbol{W}_f + \boldsymbol{b}_f)$, $\boldsymbol{i}_t=\sigma((\boldsymbol{x}_t\| \boldsymbol{h}_{t-1})\boldsymbol{W}_i + \boldsymbol{b}_i)$, and $\boldsymbol{o}_t=\sigma((\boldsymbol{x}_t\| \boldsymbol{h}_{t-1})\boldsymbol{W}_o + \boldsymbol{b}_o)$, where $\sigma$ is the sigmoid activation function and is applied entry-wise. Additionally, $\boldsymbol{\tilde{C}}_t$ represents potential information to add to the cell state: $\boldsymbol{\tilde{C}}_t=\tanh((\boldsymbol{x}_t\| \boldsymbol{h}_{t-1})\boldsymbol{W}_g + \boldsymbol{b}_g)$. Here, $\{\boldsymbol{W}_f, \boldsymbol{W}_i, \boldsymbol{W}_o, \boldsymbol{W}_g\}\in\mathbb{R}^{(m_l+ n_u)\times n_u}$ are the weights of the three gates and potential cell state update, respectively, and $\{\boldsymbol{b}_f, \boldsymbol{b}_i, \boldsymbol{b}_o, \boldsymbol{b}_g\}\in\mathbb{R}^{1\times n_u}$ are the corresponding biases.
    
    Finally, the cell state is updated based on the information from the forget gate and input gate: $\boldsymbol{C}_t = \boldsymbol{f}_t*\boldsymbol{C}_{t-1} + \boldsymbol{i}_t*\boldsymbol{\tilde{C}}_t$, and the hidden state is produced by filtering the cell state using the output gate: $\boldsymbol{h}_t=\tanh(\boldsymbol{C}_t)*\boldsymbol{o}_t$. Here again, $*$ denotes component-wise multiplication.
    
\begin{figure}
     \begin{subfigure}[scale=1]{0.5\textwidth}
         \centering
         \includegraphics[height=4.truecm]{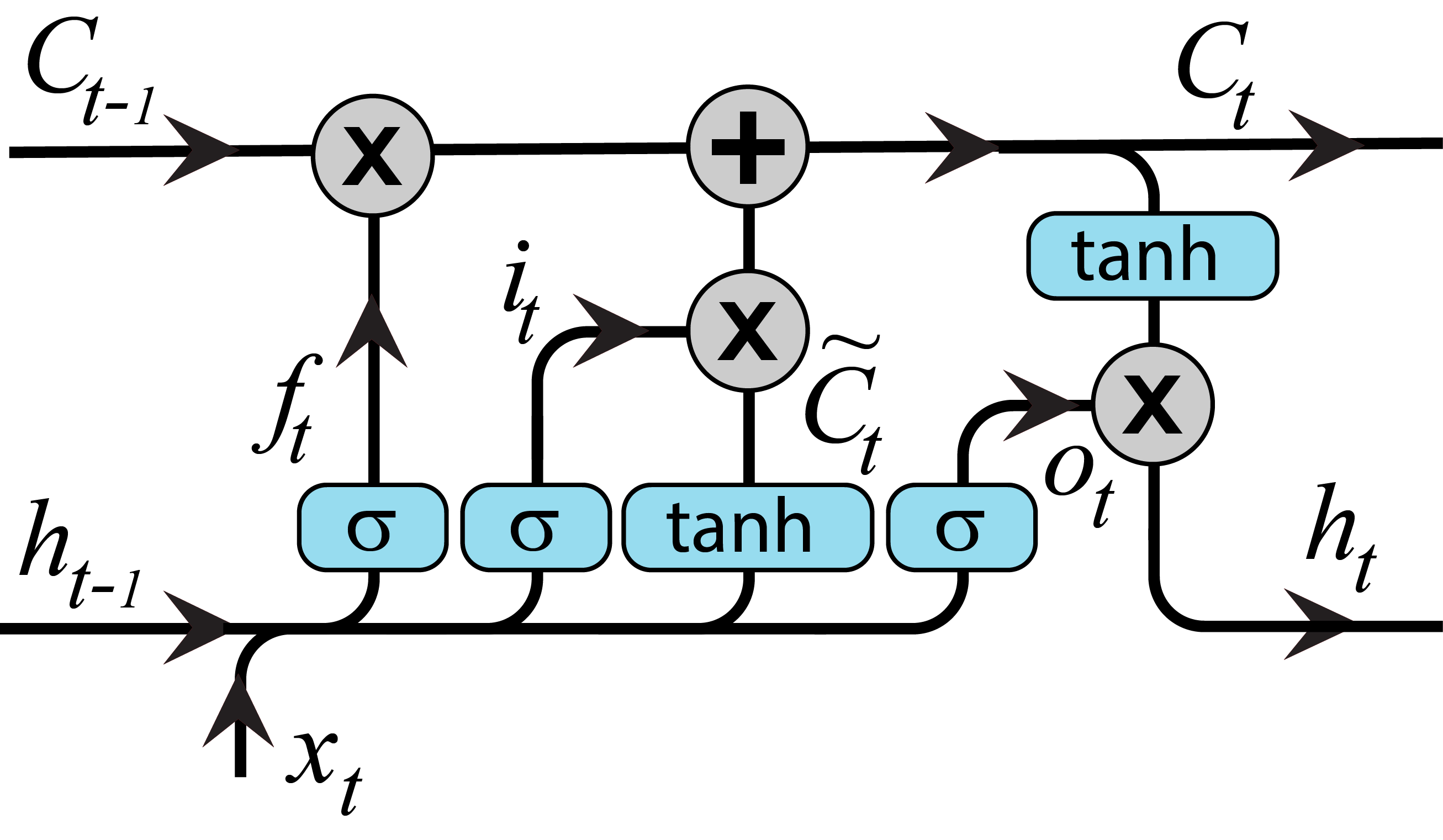}
         \caption{LSTM unit}
     \end{subfigure}
     \hfill
     \begin{subfigure}[scale=1]{0.5\textwidth}
         \centering
         \includegraphics[height=4.truecm]{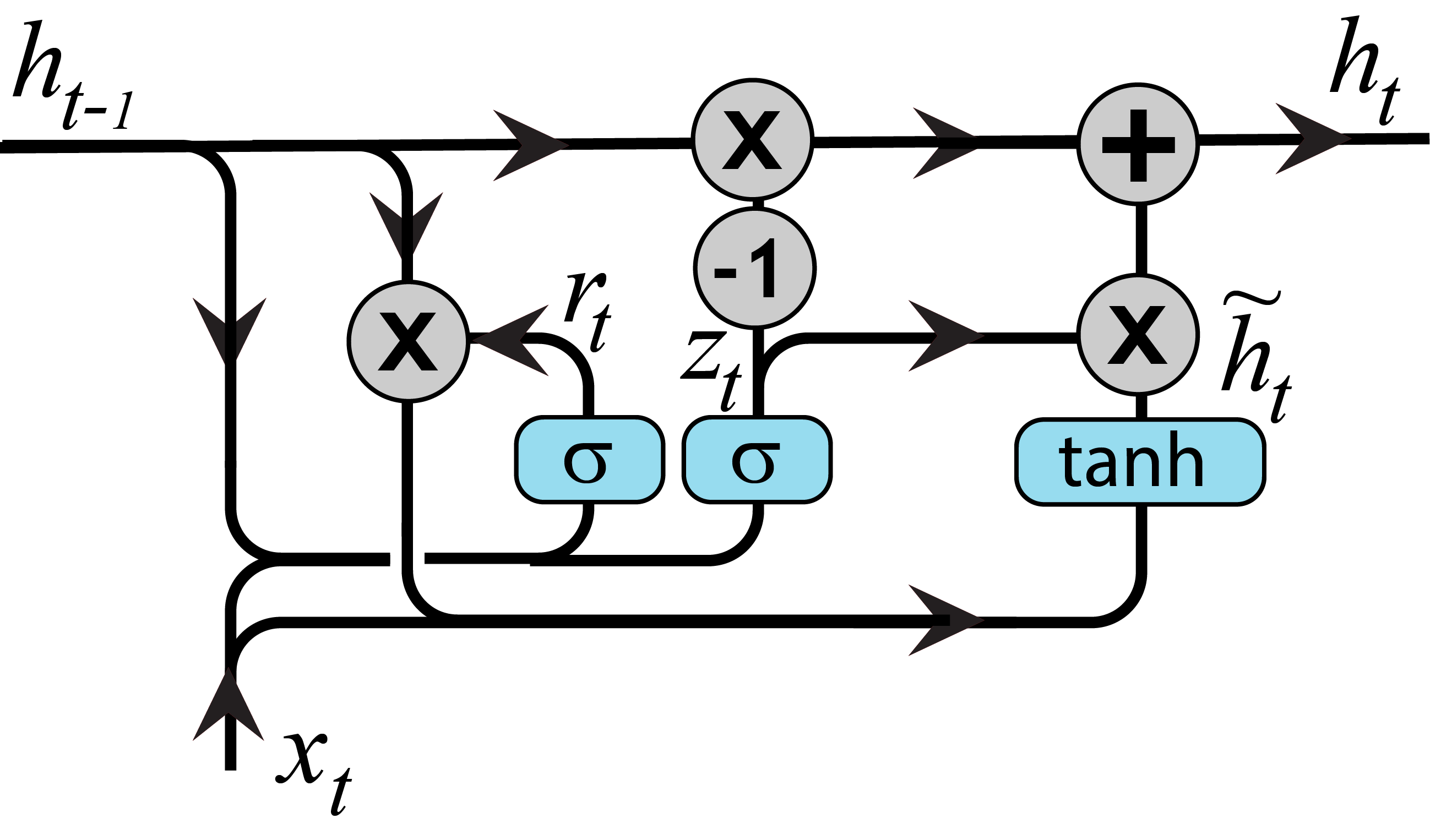}
         \caption{GRU unit}
     \end{subfigure}
        \caption{The recurrent units in the LSTM (\textit{left}) and GRU (\textit{right}) layers.}
        \label{fig:layers}
    \end{figure}

    See Figure \ref{fig:layers}a for a pictorial representation of the LSTM unit and Table \ref{tab:LSTM dims} for the input and output dimensions for the LSTM layers in \S\ref{sec:Models}. The input layer is the output of a Batch Normalization layer applied to the output of ConvLayer 2 from Table \ref{tab:FF dims}. In LSTM Layer 1, the sequential hidden states, $\boldsymbol{h}_t$, are returned for each time step $t\in[0,n_l)$, whereas in LSTM Layer 2 only the final hidden state $\boldsymbol{h}_t$ is returned, where $t=n_l-1$.
    
    \begin{table}[ht]
    \centering
    \begin{tabular}{c|l}
         \textbf{Layer} & \textbf{Input $\to$ Output Dimensions}  \\
         \hline
         LSTM Layer 1 & $[86\times 64] \to[86 \times 64]$ \\
         \hline
         LSTM Layer 2 & $[86\times 64] \to [1 \times 64]$ 
    \end{tabular}
    \caption{Input/output dimensions of the LSTM layers used in \S\ref{sec:Models}.}
    \label{tab:LSTM dims}
    \end{table}

\subsection{GRU Layers}
    
    GRU layers each have a GRU unit with hidden state length $n_u=64$ and gated structures to selectively allow information to propagate. In the GRU unit, $\boldsymbol{h}_t \in \mathbb{R}^{1\times n_u}$ represents the hidden state. The reset gate, $\boldsymbol{r}_t\in\mathbb{R}^{1\times n_u}$, controls the amount of previous information to include in the hidden state and the update gate, $\boldsymbol{z}_t\in\mathbb{R}^{1\times n_u}$, selectively adds new information to the hidden state.
    
    Mathematically, the gates are structured similarly to the LSTM gates. First, the input at time step $t$, $\boldsymbol{x}_t\in\mathbb{R}^{1\times m_l}$ is concatenated with the previous hidden state, $(\boldsymbol{x}_t\| \boldsymbol{h}_{t-1})\in\mathbb{R}^{1\times (m_l + n_u)}$. Next the gates select the information from the concatenation to be used to update the hidden state: $\boldsymbol{r}_t = \sigma((\boldsymbol{x}_t\|\boldsymbol{h}_{t-1})\boldsymbol{W}_r + \boldsymbol{b}_r)$ and $\boldsymbol{z}_t = \sigma((\boldsymbol{x}_t\| \boldsymbol{h}_{t-1})\boldsymbol{W}_z + \boldsymbol{b}_z)$ where $\{\boldsymbol{W}_r, \boldsymbol{W}_z\}\in\mathbb{R}^{(m_l + n_u)\times n_u}$ and $\{\boldsymbol{b}_r,\boldsymbol{b}_z\}\in\mathbb{R}^{1\times n_u}$ are the weights and biases of $\boldsymbol{r}_t$ and $\boldsymbol{z}_t$, respectively. 
    
    Next, $\boldsymbol{\tilde{h}}_t$ is created to represent the potential information to add to the hidden state: $\boldsymbol{\tilde{h}}_t=\tanh((\boldsymbol{x}_t\|(\boldsymbol{h}_{t-1}*\boldsymbol{r}_t))\boldsymbol{W}_h + \boldsymbol{b}_h)$ where $\boldsymbol{W}_h\in\mathbb{R}^{(m_l + n_u)\times n_u}$ and $\boldsymbol{b}_h\in\mathbb{R}^{1\times n_u}$ are the weight and biases of the potential hidden state, $\boldsymbol{\tilde{h}}_t$, and $*$ is component-wise multiplication. Finally, the hidden state is updated by weighting the potential information to add to the hidden state and the previous hidden state: $\boldsymbol{h}_t=\boldsymbol{z}_t*\boldsymbol{\tilde{h}}_t + (1-\boldsymbol{z}_t)*\boldsymbol{h}_{t-1}$. 
    
    See Figure \ref{fig:layers}b for a pictorial representation of the GRU unit and Table \ref{tab:GRU dims} for the input and output dimensions for the GRU layers in \S\ref{sec:Models}. Like the LSTM layers, the input layer is the output of a Batch Normalization layer applied to the output of ConvLayer 2 from Table \ref{tab:FF dims}. In GRU Layer 1 the sequential hidden states $\boldsymbol{h}_t$ are returned for $t\in[0,n_{x_l})$, whereas in GRU Layer 2 only the final hidden state $\boldsymbol{h}_t$ is returned for $t=n_{x_l}-1$. 
    
    \begin{table}[ht]
    \centering
    \begin{tabular}{c|l}
         \textbf{Layer} & \textbf{Input $\to$ Output Dimensions}  \\
         \hline
         GRU Layer 1 & $[86\times 64] \to [86 \times 64]$ \\
         \hline
         GRU Layer 2 & $[86 \times 64] \to [1 \times 64]$ 
    \end{tabular}
    \caption{Input/output dimensions of the GRU Layers used in \S\ref{sec:Models}.}
    \label{tab:GRU dims}
    \end{table}
    
\subsection{Fully Connected Layers and Output Layers}
    
    For fully connected (FC) layer $l$, let $\boldsymbol{x}_l\in \mathbb{R}^{1 \times (n_l\cdot m_l)}$ represent the input. Then the output of $l$, $\boldsymbol{O}^{FC}_l \in\mathbb{R}^{1\times n_{fc}}$, is: ${\boldsymbol O}^{FC}_l=\sigma(\boldsymbol{x}_l\boldsymbol{W} + \boldsymbol{b})$ for weight matrix $\boldsymbol{W}\in\mathbb{R}^{(n_l\cdot m_l) \times n_{fc}}$, bias $\boldsymbol{b}\in\mathbb{R}^{1\times n_{fc}}$, and specified output dimension $n_{fc}\in\mathbb{N^+}$, where $\sigma$ is the ReLU activation function applied entry-wise.

    In Model 1 (FF) we have two fully connected layers with $n_{fc}=64$. Table \ref{tab:FC dims} shows the input and output dimensions for the feed forward layers in \S\ref{sec:Models}.
  
    \begin{table}[ht]
        \centering
        \begin{tabular}{c|l}
             \textbf{Layer} & \textbf{Input $\to$ Output Dimensions}  \\
             \hline
             FF 1 &  $[1\times 5504] \to [1\times 64]$ \\
             \hline
             FF 2 & $[1\times 64] \to [1\times 64]$ 
        \end{tabular}
        \caption{Input/output dimensions of the FC Layers used in \S\ref{sec:Models}.}
        \label{tab:FC dims}
    \end{table}
  

    The last layer for all three models is a fully connected output layer with $n_{fc}=1$, shown in Table \ref{tab:output dims}. 
    
    \begin{table}[ht]
    \centering
    \begin{tabular}{c|l}
         \textbf{Layer} & \textbf{Input $\to$ Output Dimensions}  \\
         \hline
         FC Output & $[1\times 64] \to [1 \times 1]$ \\
    \end{tabular}
    \caption{Input/output dimensions of the FC output layers used in \S\ref{sec:Models}.}
    \label{tab:output dims}
    \end{table}

\newpage

\section{Performance Metrics}
\label{App:met_def}

\subsection{Global Performance Metrics}
\label{sec:Global Perf Metrics}
The four global fit performance metrics are defined as follows.
\begin{align*}
    R_+^2 &= \max\Bigg(0,1-\frac{\sum_{i=1}^n (y_i - \hat{y}_i)^2}{\sum_{i=1}^n (y_i - \bar{y})^2}\Bigg), \\
    \\
    NRMSE &= \frac{\sqrt{\sum_{i=1}^n \frac{1}{n} (\hat{y}_i-y_i)^2}}{y_{max} - y_{min}}, \\
    \\
    Rel.\; AUC\; Diff. &=\frac{ \sum_{i=2}^n \frac{1}{2}(y_{i-1} + y_i) - \sum_{i=2}^n \frac{1}{2}(\hat{y}_{i-1} + \hat{y}_i)}{\sum_{i=2}^n \frac{1}{2}(y_{i-1} + y_i)}, \\
    \\
    r &= \frac{\sum_{i=1}^n (y_i - \bar{y})(\hat{y_i}-\hat{y})}{\sqrt{\sum_{i=1}^n (y_i - \bar{y})^2}\sqrt{\sum_{i=1}^n(\hat{y_i}-\hat{y})^2}},
\end{align*}
where $n$ is the output sample size, $y_i$ represents the $ith$ prediction by MoLS, $\bar{y}$ is the mean prediction by MoLS over the output sample, $\hat{y}_i$ is the $ith$ prediction by the neural network model, $\hat{y}$ is the associated output sample mean, and $y_{max}-y_{min}$ is the range of the $n$ MoLS predictions. We refer the reader to Appendix \ref{app:C} for two examples of how these quantities reflect differences between abundance curves.

\subsection{Season Feature Identification: Peak Timing and Season Length}\label{sec:Feature Identification}

In this section, we assume we are given an abundance time series $\{P_k\}_{k = 1, \dots, n}$ (which could be the output of MoLS or of one of the ANNs) and describe how to identify regions where $P_k$ consistently remains above a given threshold $T$. 

We define the beginning of the range where $P_k$ remains above $T$ by finding the first day $i$ such that $P_{i}$ as well as  the following 7 days, $P_{i+1}$ to $P_{i+7}$, remain above the given threshold. We calculate a matching day $j$ to mark the return to abundance values below $T$, which is defined as the first day $j > i$ such that $P_{j}$ and abundance values in the next 7 days are lower than the threshold $T$.

Mathematically, $i$ and $j$ pairs are defined as follows:
\begin{align*}
    i &= \min_{q}\left\{q : P_{q + \ell} > T \text{ for } \ell=0, \ldots, 7\right\} \\
    j &= \min_{q}\left\{q : P_{q+\ell} < T \text{ for } \ell=0, \ldots, 7 \And q > i\right\}.
\end{align*}
It is possible that there are multiple sections throughout the year where $i$ and $j$ pairs can be generated; this is particularly true for double-peak cities in Arizona and Southern California, where the \textit{Aedes aegypti} population grows initially before dying off in the summer heat, rising again once the temperature cools. In these cases, there is a split season, and the pairs are enumerated as $(i_{k}, j_{k})_{T}$ for $k=1, \ldots, N$, where $N$ is the number of regions above the threshold. Regions with $k=2, \ldots, N$ are constrained such that $i_{k}$ must occur after $j_{k-1}$ $(j_{k-1} < i_{k} < j_{k})$.

\subsection{Season Feature Performance Metrics}
\label{sec:Feature Performance Identification}
The metrics that quantify peak timing and season lengths are based on the difference, relative to MoLS data, in onset and offset times at various points in the mosquito season for a particular year. The pairs $(i_{k}, j_{k})_{T}$ defined above for a specific threshold $T$ are calculated for both the predicted (by the ANNs) and observed (MoLS) data on a yearly basis. We use the notation $i_{m}$ and $j_{m}$ to denote onset and offset days for the observed data and $\hat{\imath}_{n}$ and $\hat{\jmath}_{n}$ to denote onset and offset days for the predicted data. Each onset day $i_{m}$ in the observed data is matched with the nearest onset day $\hat{\imath}_{n}$ in the predicted data, denoted $\hat{\imath}_{n}^{m}$, according to the minimum absolute difference of the points. This is mathematically defined as:
\begin{equation*}
    \hat{\imath}_{n}^{m} = \arg \min_{\hat{\imath}_{n}} |\hat{\imath}_{n} - i_{m}|.
\end{equation*}
Additionally, predicted days $\hat{\imath}_{n}$ are only allowed to match with a single observed day $i_{m}$. If a predicted day $\hat{\imath}_{n}$ would match with more than one observed day (i.e., there exists $m_1$, $m_2$ such that $\hat{\imath}_{n}^{m_1} = \hat{\imath}_{n}^{m_2}$), $\hat{\imath}_{n}$ is instead only matched to the day $i_{m}$ with the minimum absolute difference, and the remaining observed days are matched with the remaining predicted days. This process is similarly done for the offset days $j_{m}$ and $\hat{\jmath}_{n}$.

The resulting matchings are used to calculate onset ($D_{on}$) and offset ($D_{off}$) differences as follows
\[
D_{on}= \hat{\imath}_{n}^{m} - i_{m}, \qquad D_{off} = \hat{\jmath}_{n}^{m} - j_{m}.
\]
If $D_{on}$ is negative, the ANN abundance exceeds the threshold $T$ earlier than  MoLS ($\hat{\imath}_{n}^{m} < i_{m}$), and if $D_{on}$ is positive, the ANN abundance reaches $T$ later ($\hat{\imath}_{n}^{m} > i_{m}$). This is similarly true for $D_{off}$.

\subsection{Combined Performance Score}\label{sec:Combined Score}
For each model and testing location combination, we define a fit-based performance metric
\[
d=\left\langle\sqrt{M_1^2 +M_2^2 }\right\rangle,\quad M_1 =1-{\left(\bar{R}_+^2 \cdot \bar{r}\right)}^{1/2},\quad M_2 =\sqrt{(\overline{Rel.\; AUC\; Diff.})^2 + \overline{NRMSE}^2}
\]
and a peak and seasonal performance metric
\[
h=\frac{1}{Z}\Big[\sum_{i=1}^{N_{th}} \left(1 + {\bar p}_n (i)\right) \left(\vert \bar D_{on}^i \vert \cdot \sigma (D_{on}^i )+ \vert \bar D_{off}^i \vert \cdot \sigma (D_{off}^i )\right)^{1/2}\Big],
\]
where $\bar{*}$ and $\sigma(*)$ are the mean and standard deviation, respectively, of metric $*$ calculated over the testing years, $\bar{p}_n(i)$ is the average fraction of times the model did not reach the prescribed threshold, ${Z =\sum_{i=1}^{N_{th}} (1 + \bar{p}_n(i))}$ is a normalizing factor, and $N_{th} = 4$ is the number of thresholds. The mean $\langle \cdot \rangle$ in $d$ is calculated over locations in the testing subset. The non-negative coefficient of determination and the Pearson correlation were combined into $M_1$ as a result of the strong correlation between $R_+^2$ and $r$. The use of terms of the form $\vert \bar D \vert \cdot \sigma(D)$ in $h$ aims to penalize models with large values of $\vert \bar D_{on} \vert$ and/or $\vert \bar D_{off} \vert$ or large standard deviations. This indicates large departures from MoLS predictions, since all values of $D_{on}$ and $D_{off}$ are scaled to the average season length predicted by MoLS at the selected location. We see some instances where model predictions do not reach the threshold; this is particularly relevant for hot and dry locations at higher thresholds. To calculate $|\bar{D}|$ and $\sigma(D)$, we replace all instances where the model fails to meet a given threshold with the average of $|D|$ for all location and years at the threshold. The $1+\bar{p}_n(i)$ weight then penalizes models that do not always reach all of the prescribed thresholds.

The above information is combined into a single score $S$, which can be used to compare different models based on their performance on the testing subset:
\[
S=\sqrt{S_1^2 +S_2^2}, \qquad S_1 =\frac{d}{\max (d)}, \qquad S_2 =\frac{h}{\max (h)},
\]
where in each case, the maximum is taken over all combinations of model and location.

\newpage
\section{Seasonal feature analysis of all models}
\label{app:B}
Here we provide additional results mentioned in \S\ref{sec:Base Model Performance}, which support the conclusion that the variants improve on the base models, especially at higher threshold values.

\begin{table}[ht]
\centering
\footnotesize
\begin{tabular}{|c|c|c|c|c|c|}
    \hline
    \multirow{2}{*}{\textbf{Model}} &  \multirow{2}{*}{\textbf{Metric}} & \multicolumn{4}{c|}{\textbf{Threshold (\% of Max MoLS Prediction)}} \\
    \cline{3-6}
    & & \textbf{20\%} & \textbf{40\%} & \textbf{60\%} & \textbf{80\%} \\
	 \hline 
	 \multirow{2}{*}{FF} & $D_{on}$ & -0.006 $\pm$ 0.069 & -0.013 $\pm$ 0.067 & -0.036 $\pm$ 0.082 & -0.044 $\pm$ 0.134\\ 
	 & $D_{off}$ & 0.019 $\pm$ 0.061 & 0.022 $\pm$ 0.067 & 0.039 $\pm$ 0.103 & 0.034 $\pm$ 0.135	 \\ 

	 \hline 
	 \multirow{2}{*}{LSTM} & $D_{on}$ & 0.007 $\pm$ 0.065 & -0.005 $\pm$ 0.062 & -0.027 $\pm$ 0.07 & -0.037 $\pm$ 0.088\\ 
	 & $D_{off}$ & -0.001 $\pm$ 0.05 & 0.005 $\pm$ 0.079 & 0.01 $\pm$ 0.086 & 0.019 $\pm$ 0.088	 \\ 

	 \hline 
	 \multirow{2}{*}{GRU} & $D_{on}$ & -0.006 $\pm$ 0.057 & -0.013 $\pm$ 0.059 & -0.031 $\pm$ 0.082 & -0.035 $\pm$ 0.098\\ 
	 & $D_{off}$ & \textbf{-0.001} $\pm$ \textbf{0.044} & 0.009 $\pm$ 0.072 & 0.01 $\pm$ 0.086 & 0.011 $\pm$ 0.096	 \\ 

    \hhline{|=|=|=|=|=|=|}

    \multirow{2}{*}{FF HI} & $D_{on}$ & 0.006 $\pm$ 0.065 & -0.011 $\pm$ 0.064 & -0.027 $\pm$ 0.069 & -0.049 $\pm$ 0.105\\ 
	 & $D_{off}$ & -0.008 $\pm$ 0.046 & 0.003 $\pm$ 0.055 & 0.012 $\pm$ 0.086 & 0.004 $\pm$ 0.099\\ 

	 \hline 
	 \multirow{2}{*}{FF LO} & $D_{on}$ & 0.024 $\pm$ 0.07 & 0.007 $\pm$ 0.067 & \textbf{-0.007} $\pm$ \textbf{0.07} & -0.031 $\pm$ 0.096\\ 
	 & $D_{off}$ & -0.005 $\pm$ 0.045 & -0.01 $\pm$ 0.051 & 0.005 $\pm$ 0.075 & -0.008 $\pm$ 0.08\\ 

	 \hline 
	 \multirow{2}{*}{FF HI LO} & $D_{on}$ & \textbf{-0.001} $\pm$ \textbf{0.067} & -0.015 $\pm$ 0.063 & -0.032 $\pm$ 0.076 & -0.051 $\pm$ 0.12\\ 
	 & $D_{off}$ & -0.004 $\pm$ 0.048 & \textbf{0.002} $\pm$ \textbf{0.065} & 0.013 $\pm$ 0.075 & 0.019 $\pm$ 0.093\\ 

	 \hline 
	 \multirow{2}{*}{LSTM HI} & $D_{on}$ & -0.003 $\pm$ 0.067 & -0.018 $\pm$ 0.078 & -0.027 $\pm$ 0.083 & -0.034 $\pm$ 0.105\\ 
	 & $D_{off}$ & -0.01 $\pm$ 0.051 & -0.006 $\pm$ 0.076 & -0.001 $\pm$ 0.091 & 0.002 $\pm$ 0.092\\ 

	 \hline 
	 \multirow{2}{*}{LSTM LO} & $D_{on}$ & 0.002 $\pm$ 0.063 & -0.011 $\pm$ 0.061 & -0.02 $\pm$ 0.071 & -0.031 $\pm$ 0.109\\ 
	 & $D_{off}$ & 0.007 $\pm$ 0.06 & 0.016 $\pm$ 0.087 & 0.02 $\pm$ 0.091 & 0.023 $\pm$ 0.125\\ 

	 \hline 
	 \multirow{2}{*}{LSTM HI LO} & $D_{on}$ & 0.017 $\pm$ 0.072 & \textbf{-0.003} $\pm$ \textbf{0.083} & -0.018 $\pm$ 0.094 & -0.024 $\pm$ 0.108\\ 
	 & $D_{off}$ & -0.012 $\pm$ 0.047 & -0.008 $\pm$ 0.071 & -0.003 $\pm$ 0.1 & 0.004 $\pm$ 0.11\\ 

	 \hline 
	 \multirow{2}{*}{GRU HI} & $D_{on}$ & -0.003 $\pm$ 0.058 & -0.014 $\pm$ 0.068 & -0.022 $\pm$ 0.075 & \textbf{-0.028} $\pm$ \textbf{0.088}\\ 
	 & $D_{off}$ & -0.002 $\pm$ 0.052 & -0.006 $\pm$ 0.062 & \textbf{0.0} $\pm$ \textbf{0.059} & 0.001 $\pm$ 0.092\\ 

	 \hline 
	 \multirow{2}{*}{GRU LO} & $D_{on}$ & -0.005 $\pm$ 0.06 & -0.016 $\pm$ 0.063 & -0.031 $\pm$ 0.074 & -0.038 $\pm$ 0.098\\ 
	 & $D_{off}$ & -0.003 $\pm$ 0.044 & 0.005 $\pm$ 0.057 & 0.007 $\pm$ 0.071 & 0.016 $\pm$ 0.095\\ 

	 \hline 
	 \multirow{2}{*}{GRU HI LO} & $D_{on}$ & -0.007 $\pm$ 0.067 & -0.022 $\pm$ 0.067 & -0.038 $\pm$ 0.091 & -0.04 $\pm$ 0.092\\ 
	 & $D_{off}$ & 0.011 $\pm$ 0.05 & 0.009 $\pm$ 0.07 & 0.013 $\pm$ 0.104 & \textbf{0.0} $\pm$ \textbf{0.102}\\

    \hline 

\end{tabular}
\caption{Seasonal feature metrics for the testing subset. The double line separates the base models (top three rows) from the variant models (below). Seasonal differences for a location are scaled by the average length of the season at the 20\% threshold. Entries are formatted as $\bar D \pm \sigma(D)$ and bold entries correspond to the lowest values of $\vert \bar D \vert \cdot \sigma(D)$ for each threshold, with $D = D_{on}$ or $D_{off}$. See Appendix \ref{sec:Feature Performance Identification} for a description of the metrics.}
\label{tab:seasonal_perf_all}
\end{table}

\newpage

\section{Case Study}
\label{app:C}
\begin{figure}
\centering
\begin{subfigure}[b]{\textwidth}
   \includegraphics[width=\textwidth]{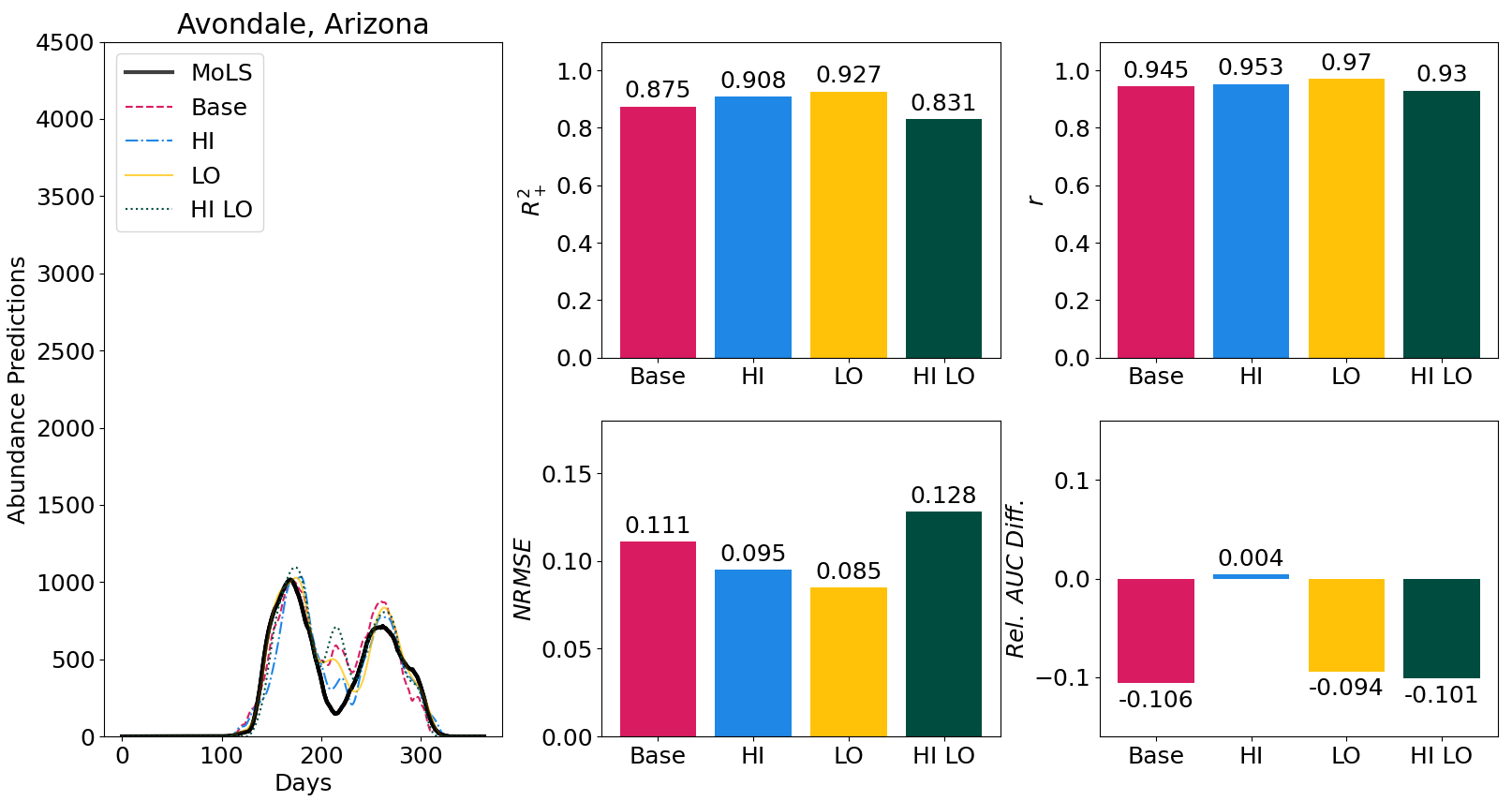}
   \caption{}
   \label{fig:avondale_preds_metrics} 
\end{subfigure}

\begin{subfigure}[b]{\textwidth}
   \includegraphics[width=\textwidth]{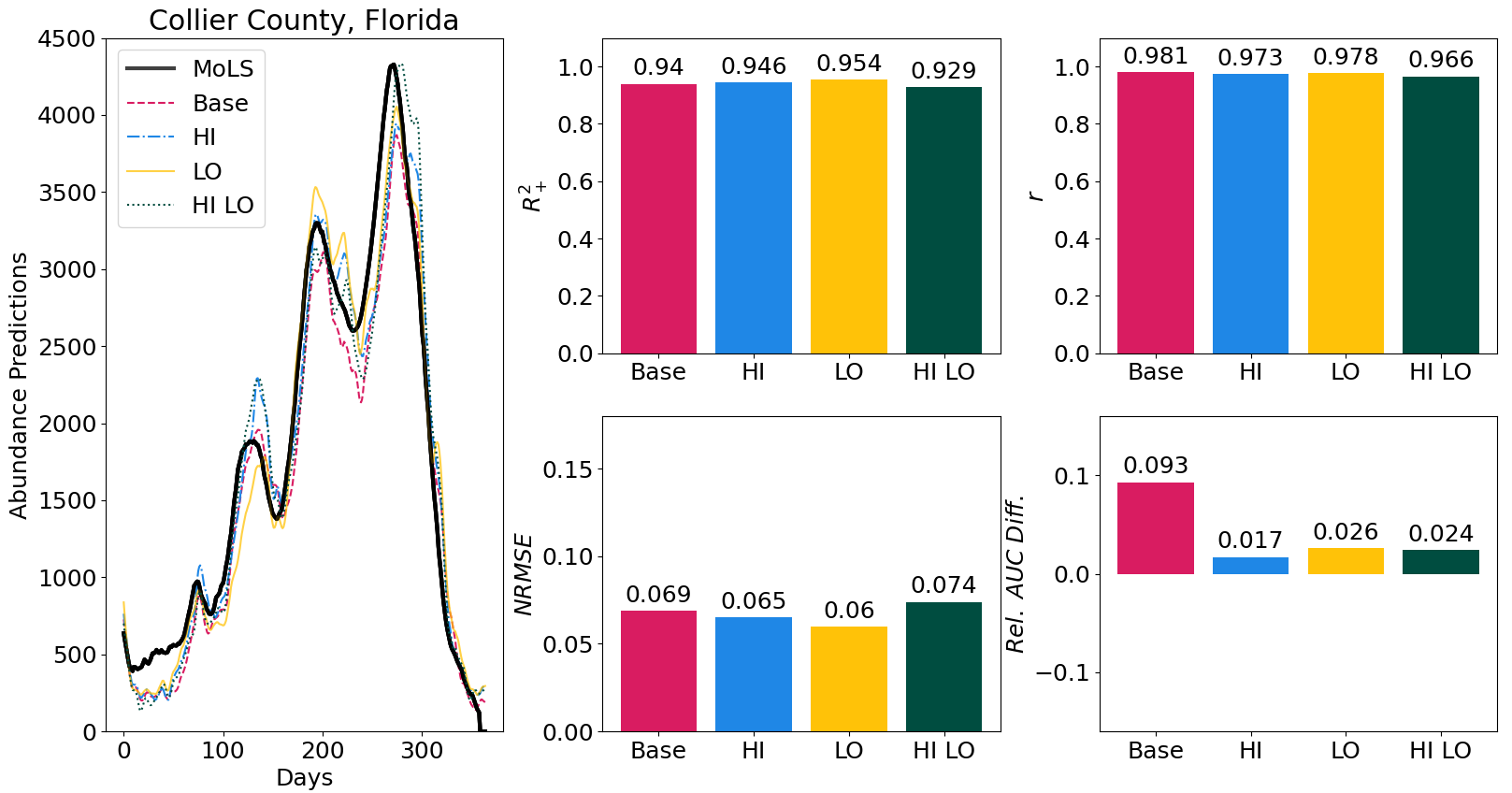}
   \caption{}
   \label{fig:collier_preds_metrics}
\end{subfigure}

\caption[Abundance Curves]{The 2020 abundance curves for the GRU model variants and associated performance metrics for (a) Avondale, Arizona and (b) Collier County, Florida. The reader is referred to \S\ref{sec:Global Perf Metrics} for descriptions of the metrics.}
\end{figure}

We present an analysis of the base GRU and GRU variant models for Avondale, Arizona and Collier County, Florida to illustrate the differences in performance observed among the model archetypes and across different climate regions. Figures \ref{fig:avondale_preds_metrics} and \ref{fig:collier_preds_metrics} compare the MoLS abundance to the estimates of the ANN models for the year 2020. Abundance curves are on the left and associated global fit metrics are on the right. Such a juxtaposition brings visual context to differences in metric values and trends observed in the main sections of the article.

In Collier County we see one long season, whereas in Avondale there are two distinct peaks. When compared to the base model, the HI and LO variants lead to improved performance in both regions, with lower $Rel.\ AUC\ Diff.$ values for the HI model and better $NRMSE$ for the LO model. In particular, the HI variant is able to capture MoLS low abundance estimates during the hot summer months in Arizona (left panel of Figure \ref{fig:avondale_preds_metrics}). On the other hand, the global fit performance of the HI LO model is comparable to or worse than that of the base model.
Tables \ref{tab:avondale_season_metrics} and \ref{tab:collier_season_metrics} display the associated seasonal metrics values, indicating slightly more consistent performance for the HI version, which achieves lowest values of $\max\left(\vert D_{on} \vert, \vert D_{off} \vert\right)$ (less than 0.048 in Avondale, Arizona and less than 0.036 in Collier County, Florida).

In summary, although all models are able to reproduce the trends observed in MoLS abundance curves in Florida, the HI version appears to better capture the dip in abundance due to hot summer months in Arizona, without deterioration of performance in other parts of the year.

\begin{table}[ht]
\small
\begin{subtable}[b]{\textwidth}
\centering
\begin{tabular}{|c|c|C{2cm}|C{2cm}|C{2cm}|C{2cm}|}
    \hline
     \multirow{2}{*}{\textbf{Model}} &  \multirow{2}{*}{\textbf{Metric}} & \multicolumn{4}{c|}{\textbf{Threshold (\% of Max MoLS Prediction)}} \\
     \cline{3-6}
     & & \textbf{20\%} & \textbf{40\%} & \textbf{60\%} & \textbf{80\%} \\
	 \hline 
	 \multirow{2}{*}{Base} & $D_{on}$ & 0.012 & -0.018 & 0.0 & -0.042\\ 
	 & $D_{off}$ & 0.024 & 0.073 & -0.012 & -0.024\\ 

	 \hline 
	 \multirow{2}{*}{HI} & $D_{on}$ & 0.0 & -0.036 & -0.048 & -0.06\\ 
	 & $D_{off}$ & 0.006 & 0.009 & -0.009 & -0.036\\ 

	 \hline 
	 \multirow{2}{*}{LO} & $D_{on}$ & 0.0 & -0.024 & -0.018 & 0.0\\ 
	 & $D_{off}$ & -0.006 & -0.057 & -0.012 & -0.024\\ 

	 \hline 
	 \multirow{2}{*}{HI LO} & $D_{on}$ & -0.03 & -0.021 & -0.021 & -0.024\\ 
	 & $D_{off}$ & 0.012 & -0.066 & -0.009 & -0.03\\ 

    \hline 
\end{tabular}
\caption{2020 Avondale, Arizona. Season length $\ell_S=165.5$ days}
\label{tab:avondale_season_metrics}
\end{subtable}

\bigskip
\begin{subtable}[b]{\textwidth}
\centering
\begin{tabular}{|c|c|C{2cm}|C{2cm}|C{2cm}|C{2cm}|}
     \hline
     \multirow{2}{*}{\textbf{Model}} &  \multirow{2}{*}{\textbf{Metric}} & \multicolumn{4}{c|}{\textbf{Threshold (\% of Max MoLS Prediction)}} \\
     \cline{3-6}
     & & \textbf{20\%} & \textbf{40\%} & \textbf{60\%} & \textbf{80\%} \\
	 \hline 
	 \multirow{2}{*}{Base} & $D_{on}$ & -0.043 & -0.023 & -0.022 & -0.047\\ 
	 & $D_{off}$ & -0.018 & -0.009 & -0.004 & 0.011\\ 

	 \hline 
	 \multirow{2}{*}{HI} & $D_{on}$ & -0.02 & -0.013 & -0.014 & -0.04\\ 
	 & $D_{off}$ & -0.016 & -0.02 & -0.014 & -0.036\\ 

	 \hline 
	 \multirow{2}{*}{LO} & $D_{on}$ & -0.051 & -0.007 & 0.004 & -0.029\\ 
	 & $D_{off}$ & -0.018 & -0.033 & -0.004 & -0.007\\ 

	 \hline 
	 \multirow{2}{*}{HI LO} & $D_{on}$ & -0.04 & -0.011 & -0.018 & -0.036\\ 
	 & $D_{off}$ & -0.011 & -0.025 & -0.014 & -0.04\\ 

	 \hline 
\end{tabular}
\caption{2020 Collier County, Florida. Season length $\ell_S=276.6$ days}
\label{tab:collier_season_metrics}
\end{subtable}

\caption{Season feature metrics for (a) Avondale, Arizona and (b) Collier County, Florida. Seasonal differences for a location and year are scaled by the average length of the season at the 20\% threshold. See \S\ref{sec:Feature Performance Identification} for a description of the metrics.}
\end{table}

\clearpage

\section{ANN Variations}
\label{App:other_models}

Table \ref{tab:base_global_var} shows global fit metrics (means and standard deviations are calculated over all years and locations in the testing subset) for the three main models and the following variations. The prime version is trained in the same way as the base version but with a different set of randomly chosen training samples. The 2000 version of each model uses 2000 instead of 1000 randomly chosen samples per training location. Finally, the $\Delta$120 version uses $\Delta = 120$ days of weather data instead of $\Delta = 90$ days. Recall that all of these models take $\Delta$ consecutive days of weather time series as input, to estimate abundance on the $\Delta^{\text{th}}$ day. Table \ref{tab:base_seasonal_var} shows similar comparisons for the seasonal metrics.

The results are highly consistent for each metric, with the mean score of each variation falling within one standard deviation of the mean scores obtained by variations of the same model. This suggests that the default choices for $\Delta$ (90 days) and for the size of the training set (1000 samples per location for the base versions), lead to models whose performance is representative of what ANNs trained on MoLS can accomplish. In addition, the variability observed in the means displayed in these tables provides intuitive context for deciding when a model performs better than another, given that each model performance is a random variable that depends on the samples randomly selected during the training process.

\begin{table}[ht]
\centering
\begin{tabular}{|c|c|c|c|c|c|}
\hline
\multicolumn{5}{|c|}{\textbf{FF Variants}}\\
\hline

\multirow{2}{*}{\textbf{Model}} & \multicolumn{4}{c|}{\textbf{Metric}} \\
\cline{2-5}
& $R_+^2$ & $NRMSE$ & $Rel.\; AUC\; Diff.$ & $r$ \\
	 \hline 
	 FF & 0.871 $\pm$ 0.142 & 0.112 $\pm$ 0.073 & 0.09 $\pm$ 0.239 & 0.961 $\pm$ 0.038\\ 

	 \hline 
	 FF$'$ & 0.901 $\pm$ 0.127 & 0.096 $\pm$ 0.074 & 0.042 $\pm$ 0.234 & 0.969 $\pm$ 0.033\\ 

	 \hline 
	 FF$_{2000}$ & 0.898 $\pm$ 0.141 & 0.095 $\pm$ 0.072 & 0.037 $\pm$ 0.223 & 0.967 $\pm$ 0.041\\ 

	 \hline 
	 FF$_{\Delta120}$ & 0.894 $\pm$ 0.14 & 0.097 $\pm$ 0.069 & 0.015 $\pm$ 0.212 & 0.966 $\pm$ 0.037\\

\hline

\multicolumn{5}{|c|}{\textbf{LSTM Variants}}\\
\hline

\multirow{2}{*}{\textbf{Model}} & \multicolumn{4}{c|}{\textbf{Metric}} \\
\cline{2-5}
& $R_+^2$ & $NRMSE$ & $Rel.\; AUC\; Diff.$ & $r$ \\
	 \hline 
	 LSTM & 0.916 $\pm$ 0.118 & 0.088 $\pm$ 0.07 & 0.031 $\pm$ 0.224 & 0.974 $\pm$ 0.028\\ 

	 \hline 
	 LSTM$'$ & 0.91 $\pm$ 0.129 & 0.089 $\pm$ 0.065 & 0.06 $\pm$ 0.196 & 0.971 $\pm$ 0.035\\ 

	 \hline 
	 LSTM$_{2000}$ & 0.919 $\pm$ 0.126 & 0.084 $\pm$ 0.071 & 0.017 $\pm$ 0.22 & 0.974 $\pm$ 0.029\\ 

	 \hline 
	 LSTM$_{\Delta120}$ & 0.904 $\pm$ 0.131 & 0.092 $\pm$ 0.075 & -0.019 $\pm$ 0.219 & 0.97 $\pm$ 0.037\\ 

\hline

\multicolumn{5}{|c|}{\textbf{GRU Variants}}\\
\hline

\multirow{2}{*}{\textbf{Model}} & \multicolumn{4}{c|}{\textbf{Metric}} \\
\cline{2-5}
& $R_+^2$ & $NRMSE$ & $Rel.\; AUC\; Diff.$ & $r$ \\
	 \hline 
	 GRU & 0.923 $\pm$ 0.119 & 0.083 $\pm$ 0.071 & 0.036 $\pm$ 0.207 & 0.975 $\pm$ 0.029\\ 

	 \hline 
	 GRU$'$ & 0.915 $\pm$ 0.146 & 0.084 $\pm$ 0.068 & -0.055 $\pm$ 0.207 & 0.975 $\pm$ 0.026\\ 

	 \hline 
	 GRU$_{2000}$ & 0.921 $\pm$ 0.118 & 0.083 $\pm$ 0.06 & 0.04 $\pm$ 0.185 & 0.975 $\pm$ 0.029\\ 

	 \hline 
	 GRU$_{\Delta120}$ & 0.921 $\pm$ 0.109 & 0.084 $\pm$ 0.052 & 0.002 $\pm$ 0.157 & 0.973 $\pm$ 0.029\\

\hline

\end{tabular}
\caption{Global performance metrics for variations of the base models on the testing subset.}
\label{tab:base_global_var}
\end{table}

\begin{table}[h!]
\small
\centering
\begin{tabular}{|c|c|c|c|c|c|}
     \hline
     \multicolumn{6}{|c|}{\textbf{FF Variants}}\\
     \hline
     \multirow{2}{*}{\textbf{Model}} &  \multirow{2}{*}{\textbf{Metric}} & \multicolumn{4}{c|}{\textbf{Threshold (\% of Max MoLS Prediction)}} \\
     \cline{3-6}
     & & \textbf{20\%} & \textbf{40\%} & \textbf{60\%} & \textbf{80\%} \\
    \hline 
	 \multirow{2}{*}{FF} & $D_{on}$ & -0.006 $\pm$ 0.069 & -0.013 $\pm$ 0.067 & -0.036 $\pm$ 0.082 & -0.044 $\pm$ 0.134\\ 
	 & $D_{off}$ & 0.019 $\pm$ 0.061 & 0.022 $\pm$ 0.067 & 0.039 $\pm$ 0.103 & 0.034 $\pm$ 0.135\\ 

	 \hline 
	 \multirow{2}{*}{FF$'$} & $D_{on}$ & 0.003 $\pm$ 0.074 & -0.011 $\pm$ 0.068 & -0.03 $\pm$ 0.079 & -0.058 $\pm$ 0.11\\ 
	 & $D_{off}$ & -0.006 $\pm$ 0.042 & 0.0 $\pm$ 0.044 & 0.018 $\pm$ 0.074 & 0.022 $\pm$ 0.089\\ 

	 \hline 
	 \multirow{2}{*}{FF$_{2000}$} & $D_{on}$ & -0.01 $\pm$ 0.064 & -0.019 $\pm$ 0.064 & -0.03 $\pm$ 0.066 & -0.049 $\pm$ 0.1\\ 
	 & $D_{off}$ & -0.005 $\pm$ 0.041 & 0.003 $\pm$ 0.057 & 0.013 $\pm$ 0.064 & 0.013 $\pm$ 0.098\\ 

	 \hline 
	 \multirow{2}{*}{FF$_{\Delta120}$} & $D_{on}$ & -0.007 $\pm$ 0.07 & -0.022 $\pm$ 0.081 & -0.032 $\pm$ 0.083 & -0.052 $\pm$ 0.107\\ 
	 & $D_{off}$ & -0.017 $\pm$ 0.041 & -0.013 $\pm$ 0.067 & -0.006 $\pm$ 0.087 & -0.008 $\pm$ 0.102\\

     \hline
     \multicolumn{6}{|c|}{\textbf{LSTM Variants}}\\
     \hline
     \multirow{2}{*}{\textbf{Model}} &  \multirow{2}{*}{\textbf{Metric}} & \multicolumn{4}{c|}{\textbf{Threshold (\% of Max MoLS Prediction)}} \\
     \cline{3-6}
     & & \textbf{20\%} & \textbf{40\%} & \textbf{60\%} & \textbf{80\%} \\
	 \hline 
	 \multirow{2}{*}{LSTM} & $D_{on}$ & 0.007 $\pm$ 0.065 & -0.005 $\pm$ 0.062 & -0.027 $\pm$ 0.07 & -0.037 $\pm$ 0.088\\ 
	 & $D_{off}$ & -0.001 $\pm$ 0.05 & 0.005 $\pm$ 0.079 & 0.01 $\pm$ 0.086 & 0.019 $\pm$ 0.088\\ 

	 \hline 
	 \multirow{2}{*}{LSTM$'$} & $D_{on}$ & -0.014 $\pm$ 0.06 & -0.021 $\pm$ 0.056 & -0.029 $\pm$ 0.076 & -0.036 $\pm$ 0.1\\ 
	 & $D_{off}$ & 0.007 $\pm$ 0.052 & 0.012 $\pm$ 0.083 & 0.021 $\pm$ 0.095 & 0.024 $\pm$ 0.114\\ 

	 \hline 
	 \multirow{2}{*}{LSTM$_{2000}$} & $D_{on}$ & 0.002 $\pm$ 0.062 & -0.01 $\pm$ 0.063 & -0.024 $\pm$ 0.086 & -0.035 $\pm$ 0.092\\ 
	 & $D_{off}$ & -0.009 $\pm$ 0.041 & -0.004 $\pm$ 0.067 & -0.001 $\pm$ 0.074 & 0.011 $\pm$ 0.089\\ 

	 \hline 
	 \multirow{2}{*}{LSTM$_{\Delta120}$} & $D_{on}$ & 0.007 $\pm$ 0.072 & -0.005 $\pm$ 0.074 & -0.015 $\pm$ 0.078 & -0.014 $\pm$ 0.098\\ 
	 & $D_{off}$ & -0.016 $\pm$ 0.05 & -0.016 $\pm$ 0.06 & -0.009 $\pm$ 0.072 & -0.0 $\pm$ 0.103\\ 

     \hline
     
    \multicolumn{6}{|c|}{\textbf{GRU Variants}}\\
     \hline
     \multirow{2}{*}{\textbf{Model}} &  \multirow{2}{*}{\textbf{Metric}} & \multicolumn{4}{c|}{\textbf{Threshold (\% of Max MoLS Prediction)}} \\
     \cline{3-6}
     & & \textbf{20\%} & \textbf{40\%} & \textbf{60\%} & \textbf{80\%} \\
	 \hline 
	 \multirow{2}{*}{GRU} & $D_{on}$ & -0.006 $\pm$ 0.057 & -0.013 $\pm$ 0.059 & -0.031 $\pm$ 0.082 & -0.035 $\pm$ 0.098\\ 
	 & $D_{off}$ & -0.001 $\pm$ 0.044 & 0.009 $\pm$ 0.072 & 0.01 $\pm$ 0.086 & 0.011 $\pm$ 0.096\\ 

	 \hline 
	 \multirow{2}{*}{GRU$'$} & $D_{on}$ & 0.042 $\pm$ 0.074 & 0.024 $\pm$ 0.073 & 0.002 $\pm$ 0.075 & -0.014 $\pm$ 0.086\\ 
	 & $D_{off}$ & -0.015 $\pm$ 0.041 & -0.02 $\pm$ 0.057 & -0.016 $\pm$ 0.051 & -0.01 $\pm$ 0.083\\ 

	 \hline 
	 \multirow{2}{*}{GRU$_{2000}$} & $D_{on}$ & -0.007 $\pm$ 0.061 & -0.015 $\pm$ 0.061 & -0.029 $\pm$ 0.073 & -0.039 $\pm$ 0.108\\ 
	 & $D_{off}$ & -0.006 $\pm$ 0.05 & -0.003 $\pm$ 0.057 & 0.006 $\pm$ 0.06 & 0.006 $\pm$ 0.098\\ 

	 \hline 
	 \multirow{2}{*}{GRU$_{\Delta120}$} & $D_{on}$ & -0.014 $\pm$ 0.059 & -0.018 $\pm$ 0.062 & -0.026 $\pm$ 0.07 & -0.024 $\pm$ 0.088\\ 
	 & $D_{off}$ & -0.009 $\pm$ 0.053 & -0.017 $\pm$ 0.05 & -0.022 $\pm$ 0.062 & -0.029 $\pm$ 0.094\\

     \hline

\end{tabular}

\caption{Seasonal fit metrics for variations of the base models on the testing subset.}
\label{tab:base_seasonal_var}
\end{table}

\clearpage

\section{Locations used in the principal data set for training, validation, and testing}
\label{app:D}

\begin{table}[ht]
\centering
\begin{filecontents*}{location_table_train.csv}
State,Training,Testing
Arizona (Cities),{Anthem*, Apache Junction*, Buckeye*, Bullhead City*, Casa Grande*, Casas Adobes, Catalina Foothills, Chandler*, Douglas, Drexel Heights, El Mirage*, Flagstaff, Florence*, Flowing Wells, Fountain Hills*, Gilbert*, Glendale*, Goodyear*, Green Valley, Kingman, Lake Havasu City*, Mesa*, New River$^{*\dagger}$, Oro Valley$^{\dagger}$, Paradise Valley*, Payson, Peoria*, Phoenix*, Prescott$^{\dagger}$, Queen Creek*, Rio Rico$^{\dagger}$, Sahuarita, San Luis*, San Tan Valley*, Scottsdale*, Sierra Vista, Somerton*, Sun City*, Surprise*, Tanque Verde, Tempe*, Yuma*},{Avondale, Eloy, Fortuna Foothills, Marana, Maricopa, Nogales, Prescott Valley, Tucson}
California (Counties),{Alameda, Contra Costa, Fresno, Glenn, Imperial*, Inyo, Kern, Kings, Lake, Los Angeles, Madera, Marin, Merced, Mono$^{\dagger}$, Monterey$^{\dagger}$, Napa$^{\dagger}$, Orange$^{\dagger}$, Placer, Sacramento$^{\dagger}$, San Benito, San Bernardino*, San Diego, San Joaquin, San Luis Obispo, San Mateo, Santa Barbara, Santa Clara, Santa Cruz, Solano, Sonoma, Sutter, Tulare},{Butte, Colusa, Riverside, Shasta, Stanislaus, Ventura, Yolo, Yuba}
Connecticut (Counties),{New Haven$^{\dagger}$},{Fairfield}
Florida (Counties),{Calhoun, Escambia, Gadsden, Hillsborough, Holmes, Jefferson, Lee, Liberty, Madison, Manatee, Martin, Miami-Dade, Okaloosa$^{\dagger}$, Pasco, Polk, Santa Rosa, St. Johns, Taylor, Wakulla, Walton, Washington},{Collier, Jackson, Osceola, Pinellas}
New Jersey (Counties),{Cumberland, Mercer, Monmouth$^{\dagger}$, Morris$^{\dagger}$, Sussex, Warren},{Essex, Salem}
New York (Counties),{Bronx$^{\dagger}$, Kings$^{\dagger}$, Nassau$^{\dagger}$, Queens, Rockland$^{\dagger}$, Westchester},{New York, Richmond}
North Carolina (Counties),{New Hanover$^{\dagger}$, Transylvania, Wake},{Forsyth, Pitt}
Texas (Counties),{Hidalgo, Tarrant},{Cameron}
Wisconsin (Counties),{Dane$^{\dagger}$, Milwaukee$^{\dagger}$},{Waukesha}
\end{filecontents*}
\csvreader[tabular=|p{0.3\textwidth}|p{0.65\textwidth}|,
    table head=\hline \textbf{State} & \textbf{Training/Validation Locations}\\\hline,
    late after line=\\\hline]
{location_table_train.csv}{State=\State,Training=\Training}
{\State & \Training}
\caption{Training and validation locations. $*$ locations are used for high  temperature oversampling and $\dagger$ are used for low temperature oversampling. States and locations are organized alphabetically.}
\label{tab:training_locs}
\end{table}

\begin{table}[ht]
\centering
\begin{filecontents*}{location_table_test.csv}
State,Testing
Florida (Counties),{Osceola, Pinellas, Collier, Jackson}
Texas (Counties),{Cameron}
New Jersey (Counties),{Salem, Essex}
New York (Counties),{New York, Richmond}
North Carolina (Counties),{Forsyth, Pitt}
Connecticut (Counties),{Fairfield}
Arizona (Cities),{Fortuna Foothills, Prescott Valley, Maricopa, Nogales, Eloy, Marana, Tucson, Avondale}
California (Counties),{Riverside, Ventura, Shasta, Yuba, Butte, Stanislaus, Yolo, Colusa}
Wisconsin (Counties),{Waukesha}
\end{filecontents*}
\csvreader[tabular=|p{0.3\textwidth}|p{0.65\textwidth}|,
    table head=\hline \textbf{State} & \textbf{Testing Locations}\\\hline,
    late after line=\\\hline]
{location_table_test.csv}{State=\State,Testing=\Testing}
{\State & \Testing}
\caption{Testing locations. States are organized bottom to top by their performance on the combined metric and locations within each state are organized left to right in the order they appear in Figure \ref{fig:combined_score}.}
\label{tab:testing_locs}
\end{table}

\newpage

\clearpage

\section{Capital Cities Data Set}
\label{app:CC}

This appendix illustrates the performance of the GRU HI model on the Capital Cities dataset. It includes 44 locations, corresponding to all capital cities in the contiguous US (48 states plus DC) situated in counties that were not used for model training, validation, or testing. Phoenix, Sacramento, Trenton, Raleigh, and Madison are therefore omitted. Results for these or nearby locations are available in the main text if they were used for testing. The weather data are the MACA time series \cite{MACA} for the Canadian Can-ESM2 \cite{Can_Model} model under the RCP4.5 scenario.
\begin{figure}[h!]
    \centering
    \includegraphics[width=\textwidth]{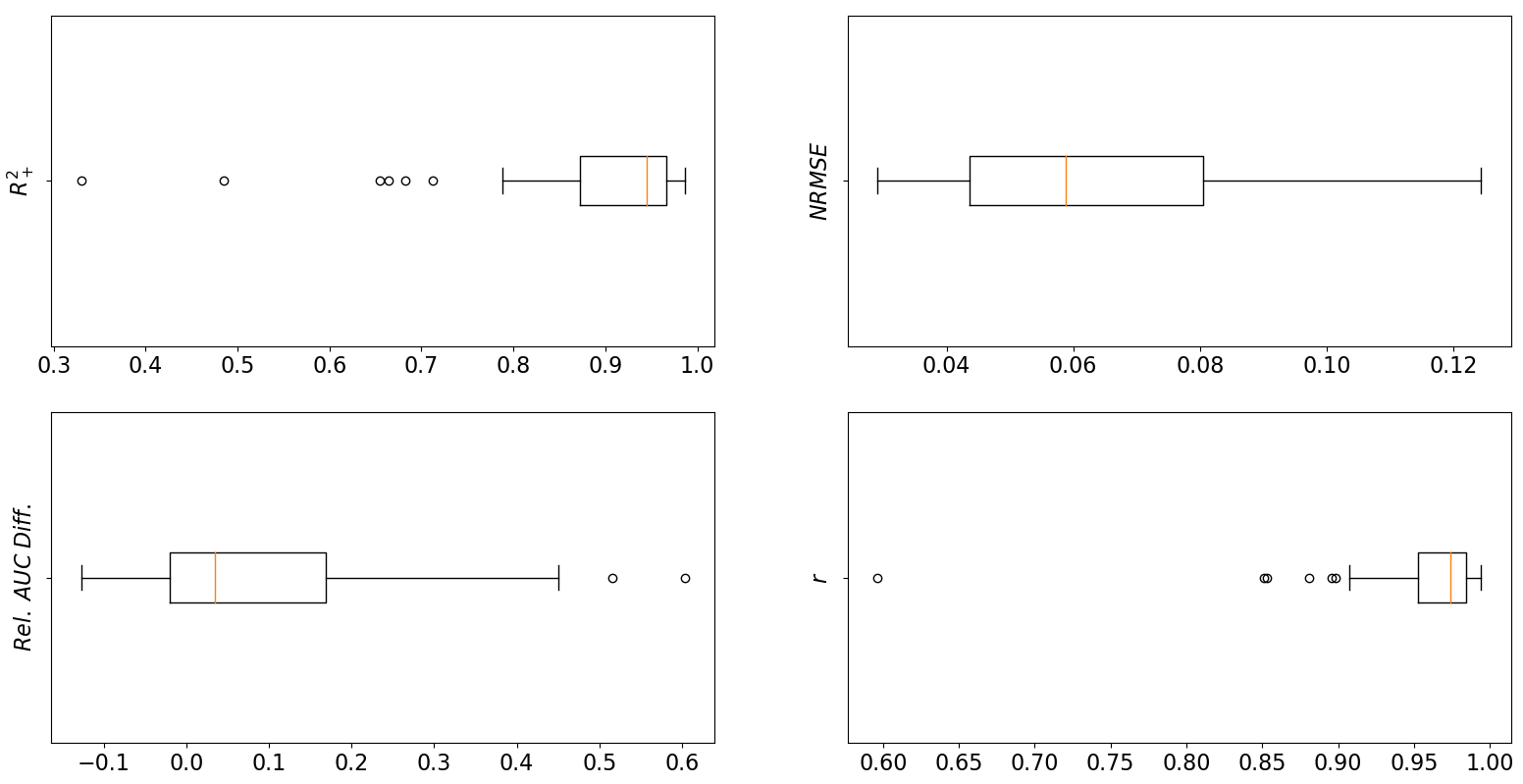}
    \caption{Global performance metrics for the GRU HI model on the Capital Cities data set. The metrics are calculated over all years, so each data point represents the metric for one capital city from 2012-2020.}
    \label{fig:global_box_plots}
\end{figure}

Figure \ref{fig:global_box_plots} displays box plots for the four global metrics assessing how the ANN estimates compare to those of MoLS, over nine years (2012-2020) of weather data. We see that the GRU HI model replicates MoLS time series very well, both in trend and  size, except for a few outliers listed in Table \ref{tab:global_box_plots_outliers}.

\begin{table}[h!]
    \centering
    \begin{tabular}{|c|l|}
    \hline
    \textbf{Metric} & \textbf{Outliers} \\
    \hline
         $R^2$ & Idaho, Nevada, North Dakota, Oregon, Vermont, Wyoming  \\
         \hline
         $Rel.\;AUC\;Diff.$ & Oregon, Wyoming \\
         \hline
         $r$ & Idaho, Nevada, North Dakota, Oregon, Vermont, Wyoming \\
         \hline
    \end{tabular}
    \caption{Capital city outliers in the global performance box plots, Figure \ref{fig:global_box_plots}.}
    \label{tab:global_box_plots_outliers}
\end{table}

\begin{figure}[ht]
    \centering
    \includegraphics[width=\textwidth]{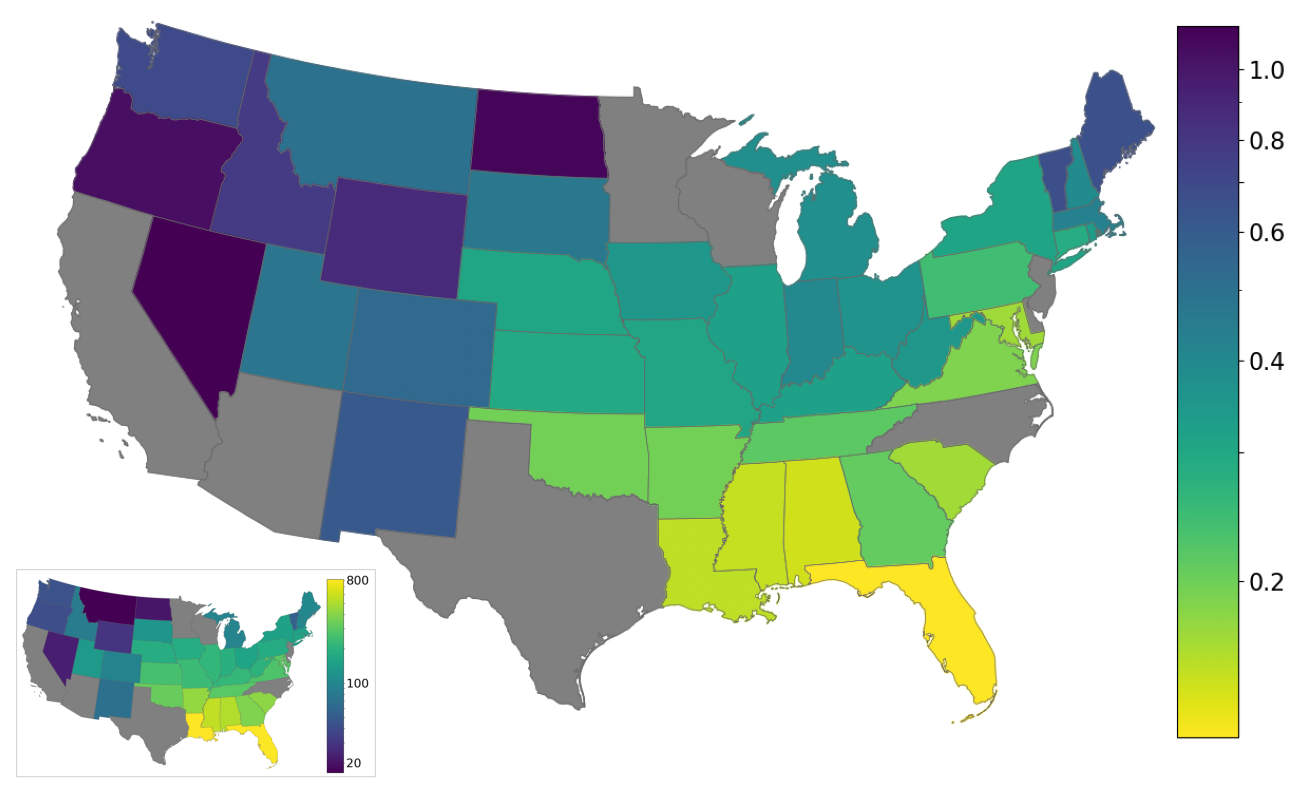}
    \caption{Combined scores on capital cities (see \S\ref{sec:Combined Score} for metric definition). Each state is colored according to the corresponding capital city score. The inset shows the average daily number of mosquitoes predicted by MoLS for each capital city; the log scale ranges from 16.44 (dark purple) to 814.28 (yellow). States in gray were omitted since the capital city was part of either the training or the testing subsets. We see a strong correlation between score and number of mosquitoes.}
    \label{fig:capitals_score_map}
\end{figure}

The $D_{on}$ and $D_{off}$ metrics for seasonal fit show good performance as well, but with a larger number of outliers (see Table \ref{tab:seasonal_box_plots_outliers}), which now include Arkansas, Colorado, Indiana, Maine, Montana, New Hamshire, New Mexico, Utah, and Washington State. The number of outliers decreases as the threshold increases, suggesting the GRU HI model does not reach the $80\%$ threshold in some of the poor performing locations. This is captured in the combined score, Figure \ref{fig:capitals_score_map}.
\begin{figure}[h!]
    \centering
    \includegraphics[width=\textwidth]{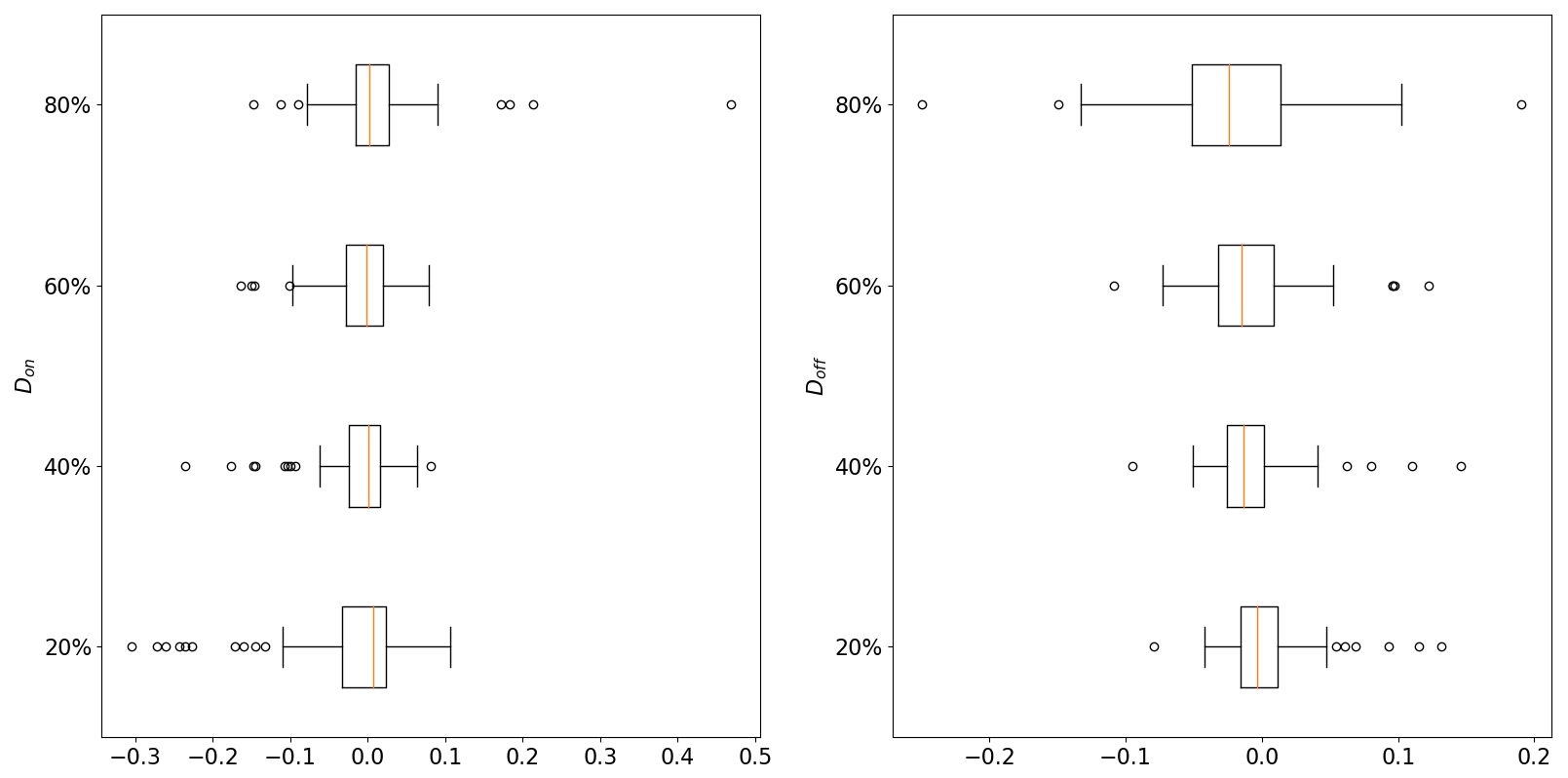}
    \caption{Seasonal fit metrics for the GRU HI model on the Capital Cities data set. The metrics are averaged over all years, so each data point represents the average value for one capital city.}
    \label{fig:seasonal_box_plots}
\end{figure}

\begin{table}[h!]
    \centering
    \begin{tabular}{|c|c|l|}
    \hline
    \textbf{Metric} & \textbf{Threshold} & \textbf{Outliers} \\
    \hline
    \multirow{7}{*}{$D_{on}$} & \multirow{2}{*}{$20\%$} & Maine, Montana, Nevada, New Hampshire, New Mexico, \\
    & & North Dakota, Oregon, Vermont, Washington, Wyoming \\
    \cline{2-3}
    & \multirow{2}{*}{$40\%$} & Idaho, Indiana, Maine, Montana, New Hampshire, \\
    & & New Mexico, Oregon, Vermont, Wyoming \\
    \cline{2-3}
    & $60\%$ & Colorado, Maine, New Mexico, Wyoming \\
    \cline{2-3}
    & \multirow{2}{*}{$80\%$} & Arkansas, Maine, Montana, Nevada, North Dakota, \\
    & & Oregon, Washington \\
    \hline
    \multirow{5}{*}{$D_{off}$} & \multirow{2}{*}{$20\%$} & Idaho, Montana, Nevada, North Dakota, Oregon, \\
    & & Utah, Wyoming \\
    \cline{2-3}
    & $40\%$ & Idaho, Nevada, Oregon, Vermont, Wyoming \\
    \cline{2-3}
    & $60\%$ & Idaho, New Mexico, Oregon, Utah, Vermont \\
    \cline{2-3}
    & $80\%$ & North Dakota, Oregon, Vermont \\
    \hline
    \end{tabular}
    \caption{Capital city outliers in the seasonal performance box plots, Figure \ref{fig:seasonal_box_plots}.}
    \label{tab:seasonal_box_plots_outliers}
\end{table}

We observe a strong correlation between the ANN performance and the number of mosquitoes predicted by MoLS: performance is lower in regions where expected mosquito numbers are lower overall. This is illustrated in Figure \ref{fig:capitals_score_map}. The main map shows the combined score (defined in \S\ref{sec:Combined Score}) of the GRU HI model. Each state is colored according to the performance of the ANN on weather data for its capital city, with yellow indicating high performance (low score) and purple low performance (score of $1$ or higher). States colored in gray are those not included in the Capital Cities dataset. The inset shows 
a similar map, but now with colors indicating MoLS average abundance for the years in the dataset (2012-2020). Dark colors correspond to very low average abundance (less than 20 mosquitoes per day) while light colors indicate average abundance over 800 mosquitoes per day. Both scales are logarithmic. The log scale correlation (Pearson coefficient $r = -0.913$; data not shown) is evident visually and supports the conclusion that the GRU HI model is a good MoLS replacement in regions of high mosquito abundance.

\begin{figure}[h!]
    \centering
    \includegraphics[width=\textwidth]{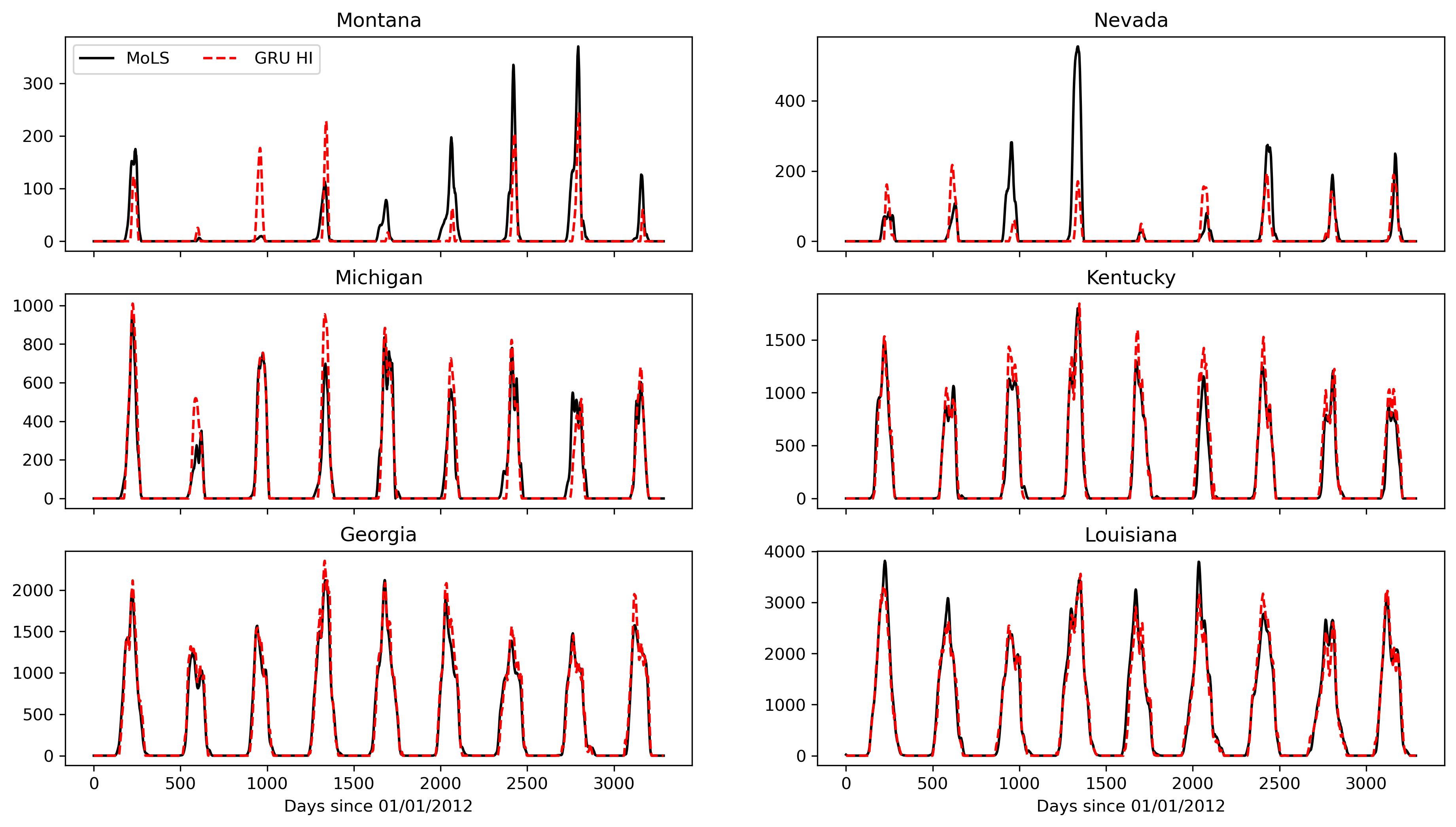}
    \caption{Comparison of the GRU HI output to MoLS data for capital cities of varying scores (Figure \ref{fig:capitals_score_map}).}
    \label{fig:GRUHI_MoLS_Comparison}
\end{figure}

To illustrate how good and bad performance scores translate into errors on mosquito estimates, Figure \ref{fig:GRUHI_MoLS_Comparison} plots the abundance time series generated by MoLS (black solid curves) and by the GRU HI model (red dashed curves) in locations of low (Nevada and Montana), medium (Kentucky and Michigan), and high (Georgia and Louisiana) performance based on their combined scores.

\clearpage

\newpage

\section*{Supplementary Material - MoLS-GRU HI Comparison}
The figures below show how the output of the GRU HI model compares to MoLS data over a period of 9 years (2012-2020). Each panel corresponds to one column of Figure \ref{fig:combined_score} and the locations are in the same order as in Figure \ref{fig:combined_score}.

\begin{center}
\includegraphics[width=\textwidth]{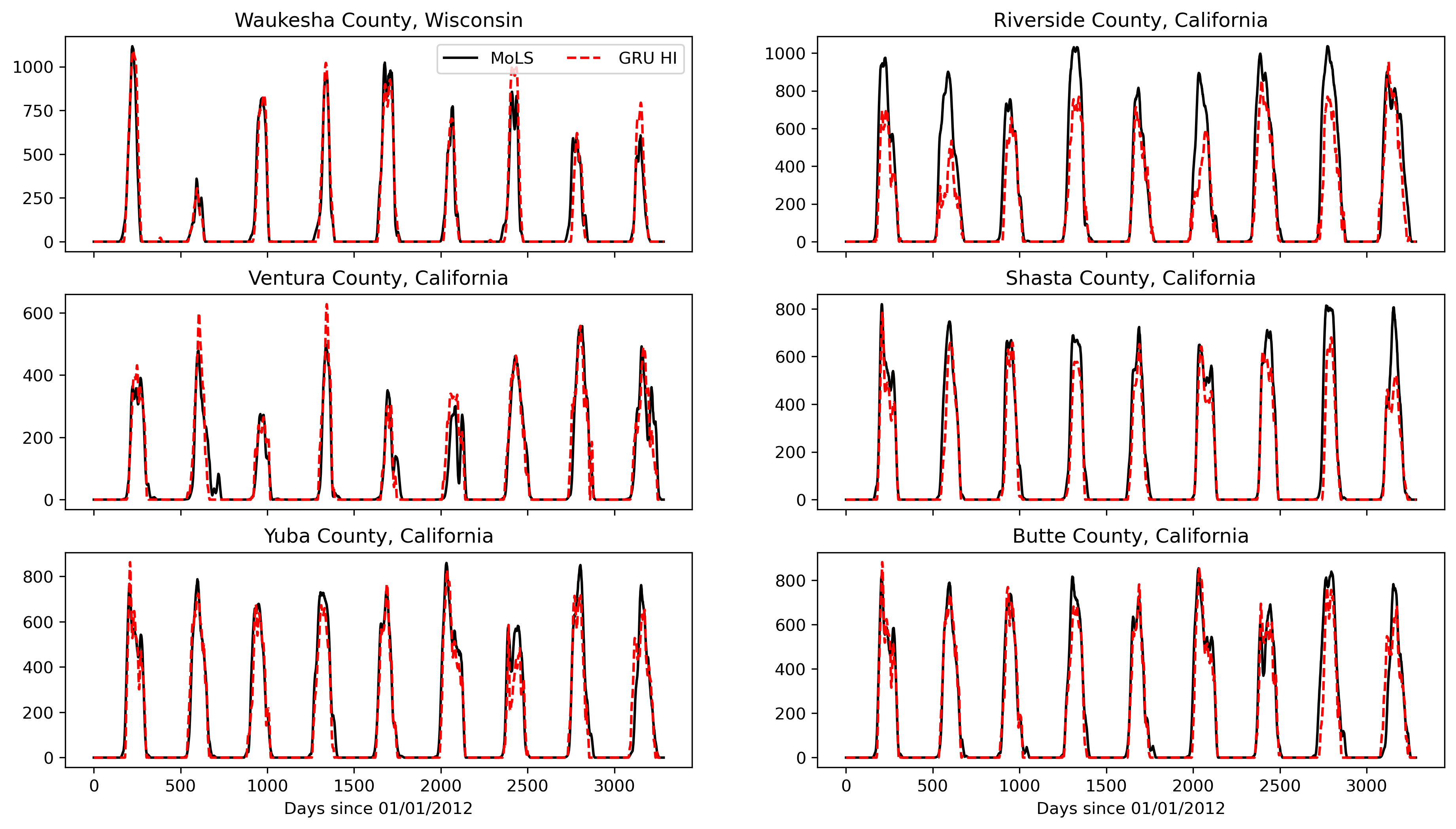}\\
\includegraphics[width=\textwidth]{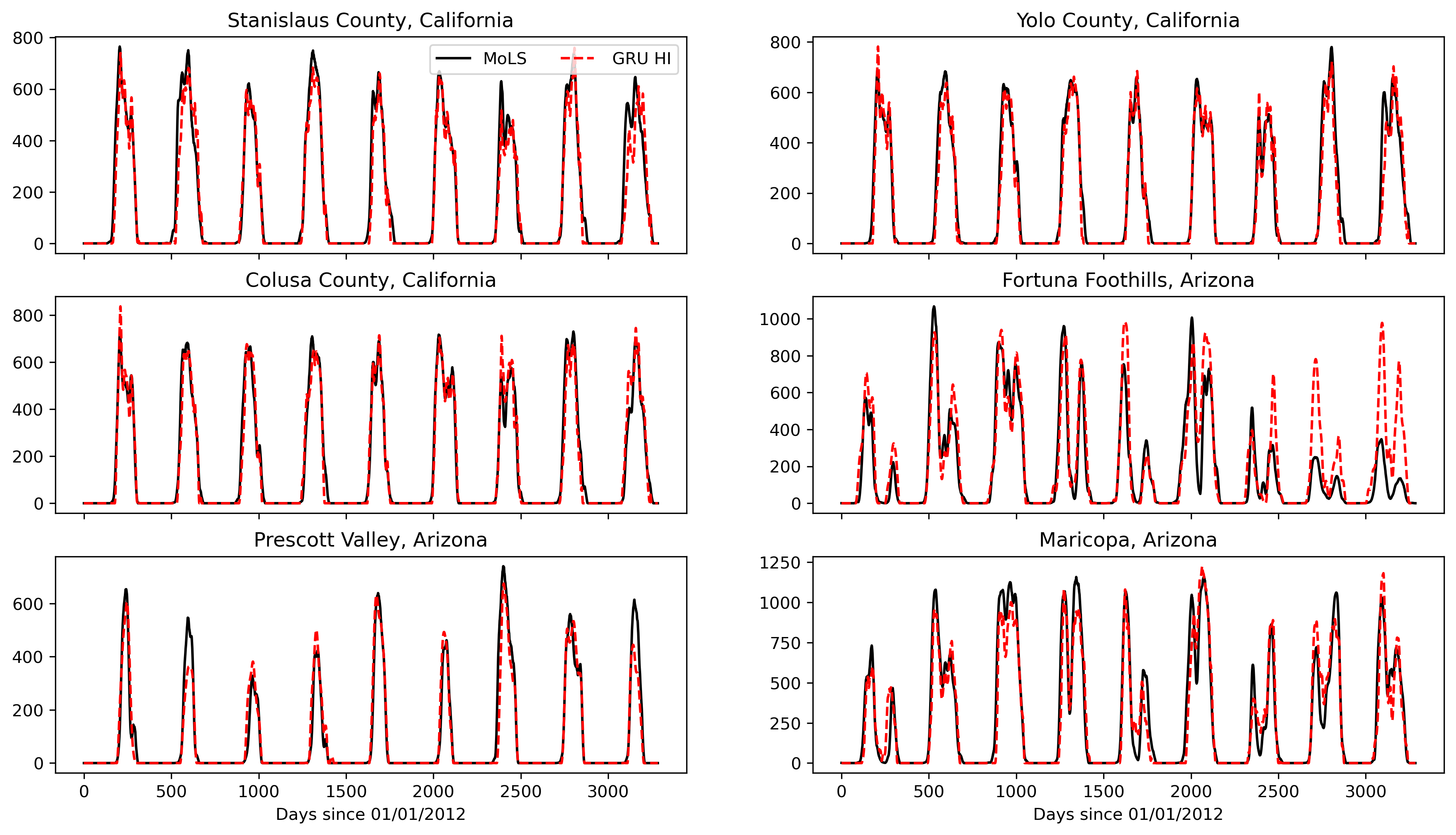}\\
\includegraphics[width=\textwidth]{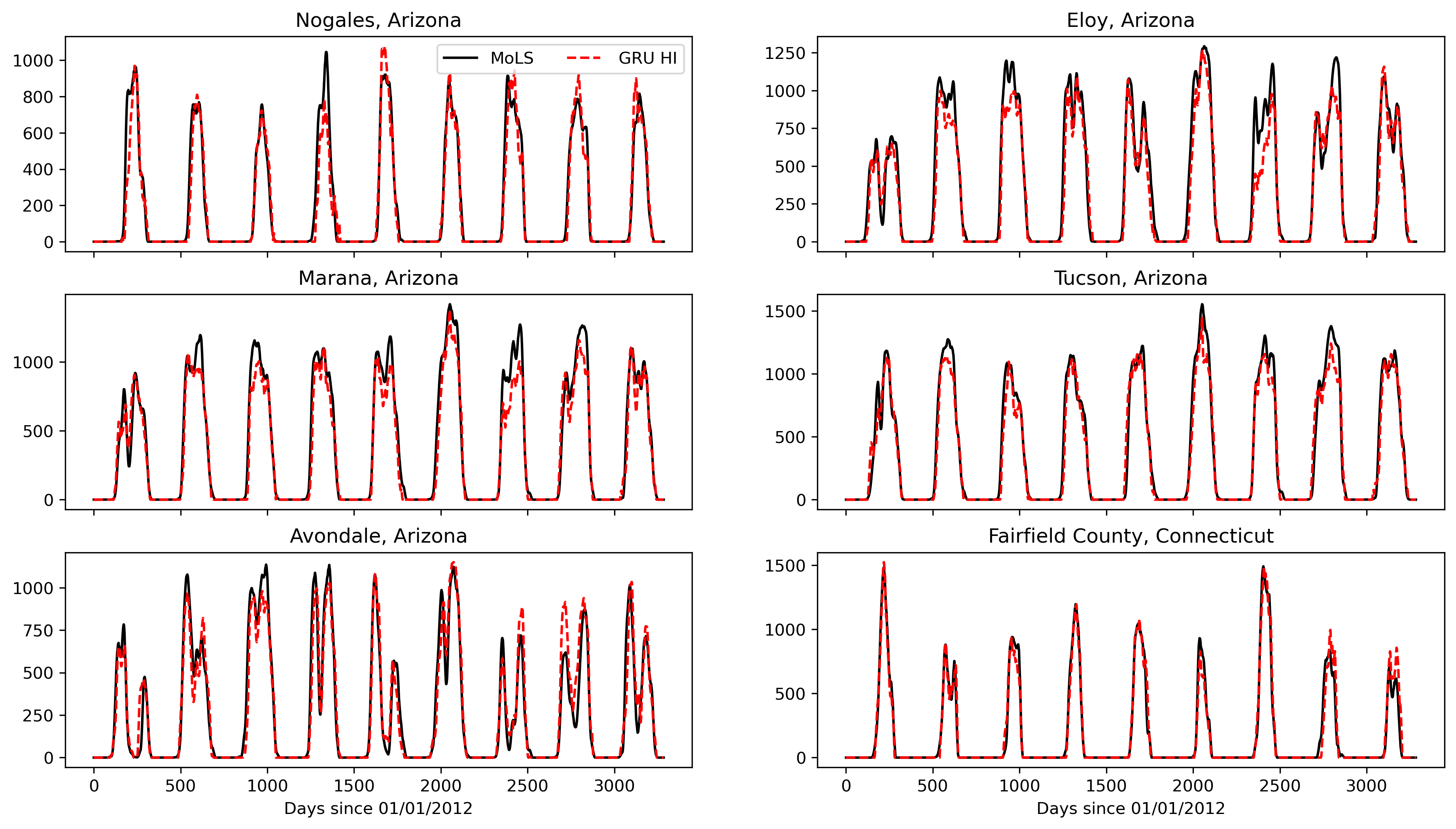}\\
\includegraphics[width=\textwidth]{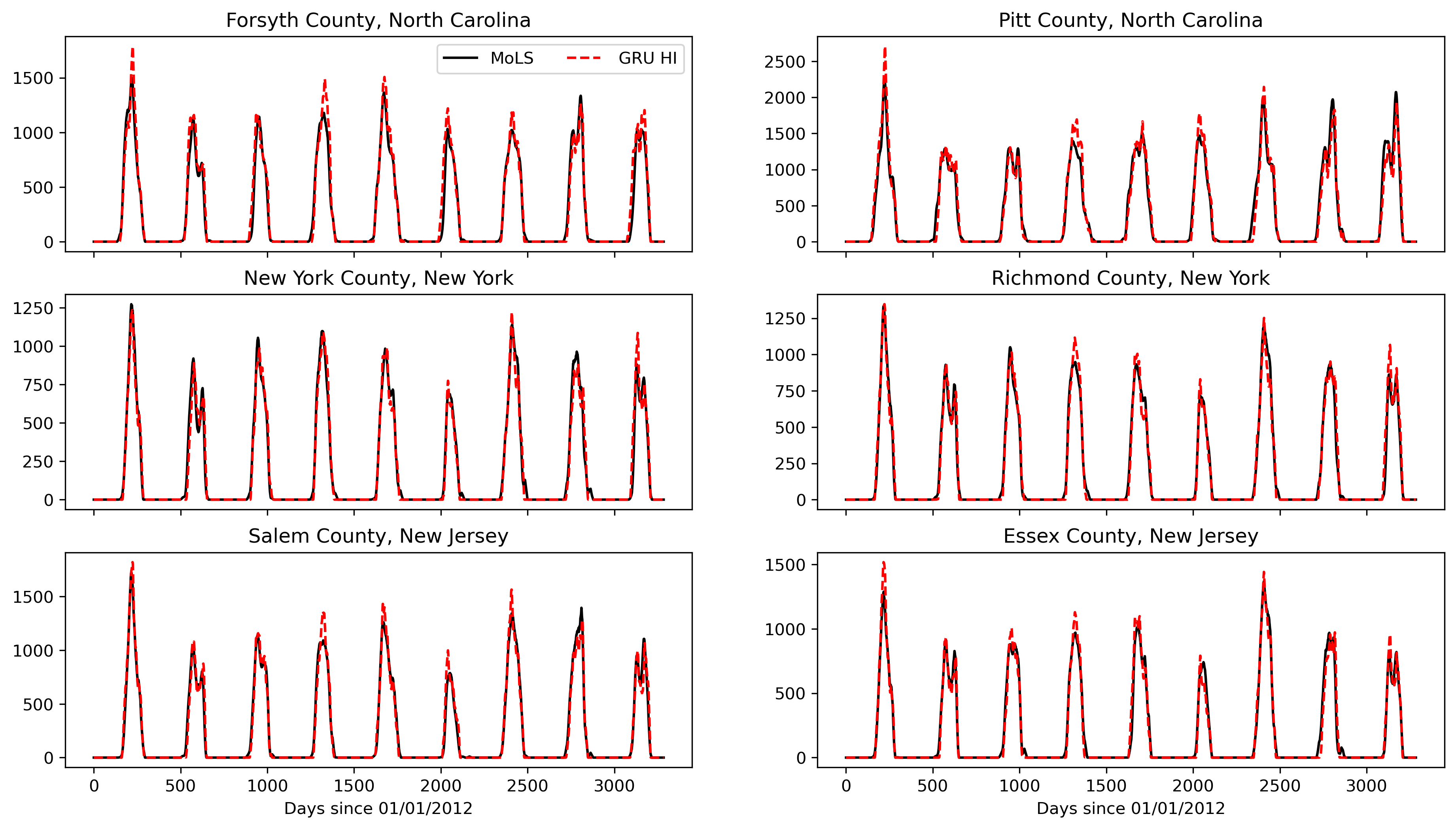}\\
\includegraphics[width=\textwidth]{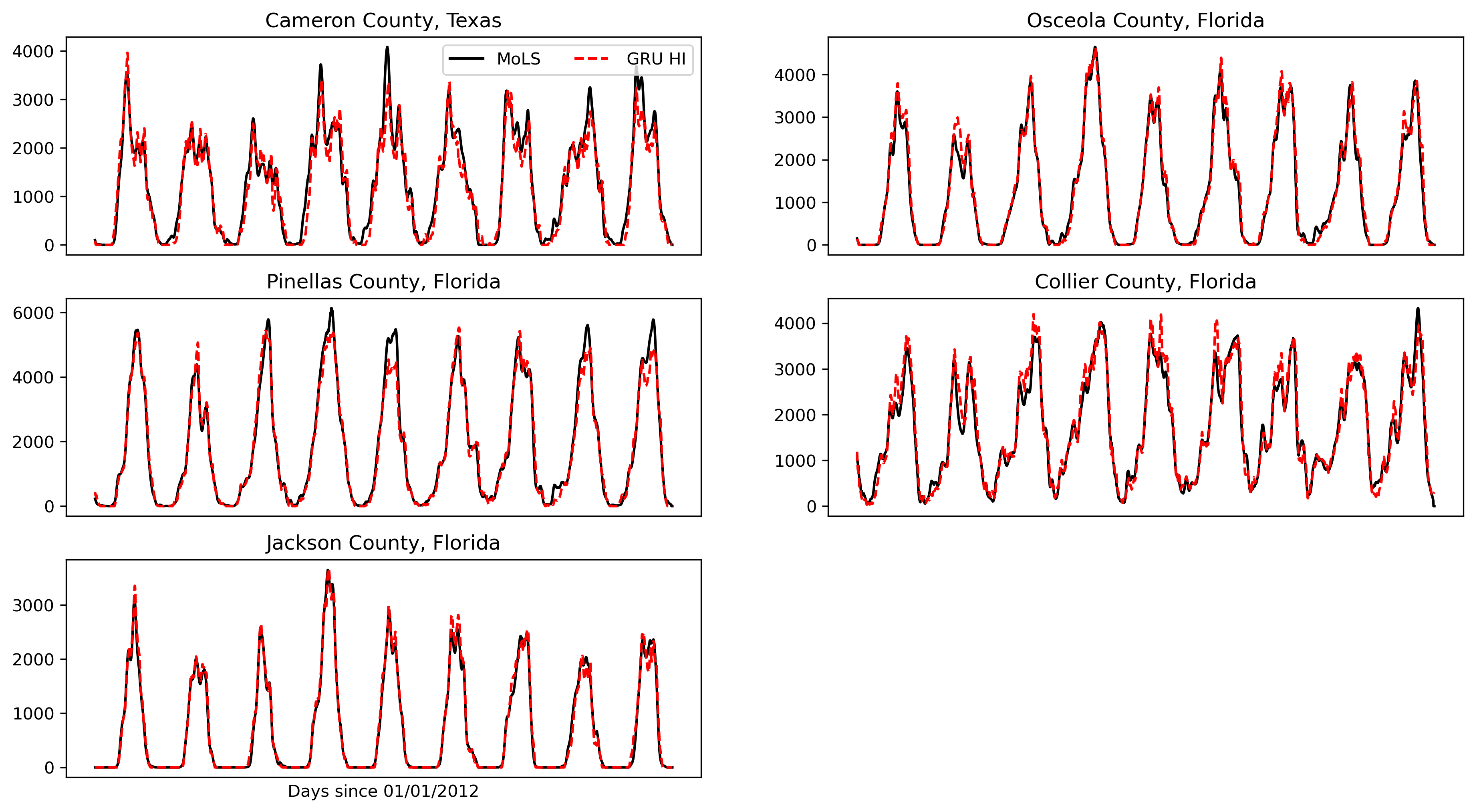}
\end{center}

\end{document}